\documentstyle[epsfig,12pt,preprint,tighten,aps]{revtex}

\begin{document}
\begin{titlepage}
\rightline{{\tt May 2001}}

\vskip 1.8 cm

\centerline{{\large \bf Solutions of the atmospheric, solar and
LSND neutrino anomalies}}

\vskip 0.8 cm

\centerline{{\large \bf from TeV scale quark-lepton unification}}

\vskip 1.3 cm

\centerline{T. L. Yoon and R. Foot}

\vskip 1.0 cm\noindent

\centerline{{\it School of Physics}}

\centerline{{\it Research Centre for High Energy Physics}}

\centerline{{\it The University of Melbourne}}

\centerline{{\it Victoria
 3010 Australia }}

\vskip 2.0cm

\centerline{Abstract}

There is a unique $SU(4) \otimes SU(2)_L \otimes SU(2)_R$ gauge
model which allows quarks and leptons to be unified at the TeV
scale. It is already known that the neutrino masses arise
radiatively in the model and are naturally light. We study the
atmospheric, solar and LSND neutrino anomalies within the
framework of this model.

\vskip 0.7cm

\noindent

\end{titlepage}

\section{Introduction}
The similarity of the quarks and leptons may be due to some type
of symmetry between them. Various theories have been proposed
including the Pati-Salam theory \cite {ps} where the leptons take
the fourth colour within a $SU(4)\otimes SU(2)_L \otimes SU(2)_R$
gauge model\footnote{ Other possibilities include models with a
discrete quark-lepton symmetry which features a spontaneously
broken $SU(3)$ colour group for leptons.\cite{QLS}}. While
definitely a good idea, there are some slightly unpleasant aspects
of the Pati-Salam model. In particular, one of the main drawbacks
of the Pati-Salam theory is that almost all of its unique
predictions for new physics cannot be tested because the
experimental constraints on the symmetry breaking scale imply that
it is out of reach of current and proposed experiments. The
problem is twofold. First there are stringent constraints coming
from rare meson decays. These imply a lower limit on the symmetry
breaking scale of about $20\ $TeV\cite{willen} for the symmetry
breaking scale which means that the heavy gauge bosons are too
heavy to be found in even the Large Hadron Collider (LHC). The
Pati-Salam model at the relatively low scale of 20 TeV also has
great problems in explaining the light neutrino masses. The
see-saw mechanism adopted in such models cannot suppress the
neutrino mass sufficiently unless the symmetry breaking scale is
very high. The required light neutrino masses suggest that the
symmetry breaking scale is at least about 50 PeV \cite{nir} (1 PeV
$\equiv $ 1000 TeV). All is not lost however. There appears to be
a unique alternative  $SU(4)\otimes SU(2)_L \otimes SU(2)_R$ gauge
model (which we call the alternative 422 model) which preserves
the elegance and simplicity of the original Pati-Salam theory
while allowing for a low symmetry breaking scale of about 1
TeV\cite {422.1}. This not only allows the unique predictions of
the theory to be tested in collider experiments, but the theory
also avoids the dreaded gauge hierarchy problem by not introducing
any hierarchy to begin with. The theory also has characteristic
predictions for rare $B, K$ decays, baryon number violation as
well as non-zero neutrino masses, all of which are naturally
within current bounds, despite the low symmetry breaking scale of
a TeV. Thus in one act of prestidigitation all of the problems
aflicting the original Pati-Salam model are cured.

Over the last few years, significant evidence for neutrino masses
has emerged from the neutrino physics anomalies: the atmospheric
\cite{atm}, solar \cite{solar} and LSND \cite{LSND} neutrino
experiments. It is therefore an interesting question as to whether
the alternative 422 model can accommodate these experiments. It
has already been shown \cite{422.2} that the masses for the
neutrinos in this model typically span the necessary range to
possibly account for these experiments. In fact viewed simply as a
gauge model for neutrino masses, the theory is quite interesting
because it provides a nice explanation for the small masses of the
neutrinos without any need for (untestable) hypothesis about high
energy scales (which arise in most popular theories of neutrino
masses). We will show that the theory in its minimal form can
accommodate the atmospheric and LSND neutrino anomalies but not
all three (including solar) simultaneously. Thus, the theory is a
candidate for the physics responsible for the neutrino physics
anomalies. However, since it cannot explain all three of the
neutrino anomalies, it obviously follows that if all three
anomalies are confirmed in forthcoming experiments, then this
would require physics beyond this model for an explanation. One
elegant possibility is the mirror symmetrized extension which can
provide a simple explanation of the neutrino physics anomalies, as
we will show.

The outline of this paper is as follows: In section II we briefly
comment on the experimental situation and the various oscillation
solutions to the solar, atmospheric neutrino anomalies and the
LSND measurements. In section III we revise the essentials of the
alternative 422 model. In particular we will derive the mass
matrices of the fermions after spontaneous symmetry breaking
(SSB). We will also define a basis for the fermions in which their
weak eigenstates are related to their respective mass eigenstates
via CKM-type unitary matrices. In section IV we explain the
mechanisms that give rise to the masses of the neutrinos. This
includes the tree level mixing that generates right-handed
neutrino Majorana masses 
\footnote{Throughout this paper the tilde on the fermion
fields is to signify that they are flavour eigenstates as opposed to
mass eigenstates (without a tilde) in this
model. See Section III for more detail. Note that the tilde has
nothing to do with supersymmetry other than a mere notation
coincidence.} 
$\overline{\tilde \nu_R} M_R (\tilde \nu_R)^c$
and two radiative mechanisms, via gauge and scalar
interactions, that give rise to Dirac masses of the ordinary
neutrinos. In section V we consider the special case of decoupled
generations and examine the possibility to obtain near maximal
$\tilde \nu_L \to (\tilde \nu_R)^c$ oscillations within the model
(which turns out to be negative). In section VI, we mirror
symmetrise the alternative 422 model to obtain a TeV scale
solution scheme for all three of the neutrino anomalies. In
section VII we will identify how the minimal alternative 422 model
could provide simultaneous solutions to (near maximal
oscillations) atmospheric neutrino anomaly and the LSND
measurements in the case where gauge interactions dominates the
radiative neutrino mass generation. We conclude in section VIII.

\section{Oscillation solutions to the atmospheric and solar neutrino
anomalies and LSND measurements}

The experimental situation regarding the oscillation solutions of
the atmospheric and solar neutrino problems has made much progress
over the last few years. In this paper we mainly focus on the
simplest possible solutions which involve maximal (or near
maximal) two-flavour oscillations.

\subsection{Atmospheric neutrino anomaly}

In the case of atmospheric neutrino anomaly there is compelling
evidence that about half of the up-going $\nu_\mu$ flux disappears
\cite{atm}. The simplest oscillation solutions which can explain
the data are maximal $\nu_\mu \to \nu_\tau$ or maximal $\nu_\mu
\to \nu_{sterile}$ oscillations \cite{FVYa}. Despite impressive
efforts by Super-Kamiokande \cite{Fukuda} the experimental data
cannot yet distinguish between these two possibilities
\cite{Fexclude}. Unfortunately, this situation probably cannot be
clarified until long-baseline experiments provide or fail to
provide $\tau$ events approximately in the year 2007. At present
the parameter range that is consistent with the atmospheric
neutrino anomaly is roughly $\sin^2 2\theta \stackrel{>}{\sim}
0.85$ and
\begin{eqnarray}
10^{-3} \stackrel{<}{\sim} \delta m^2_{atm}/\hbox{eV}^2
\stackrel{<}{\sim} 10^{-2}.
\end{eqnarray}

\subsection{Solar neutrino anomaly}

In the case of solar neutrino anomaly \cite{solar} there is very
strong evidence that about half of the $\nu_e$ flux from the sun
has gone missing when compared to theoretical models. The simplest
explanation of this is in terms of maximal $\nu_e$ oscillations,
with the main suspects being maximal $\nu_e \to \nu_\alpha$
($\nu_\alpha$ is some linear combination of $\nu_\mu$ or
$\nu_\tau$)\cite{conf,guth,flv,sg} or maximal $\nu_e \to
\nu_{sterile}$ oscillations \cite{flv,sg,cfv} \footnote{Note that
maximal (or near maximal) $\nu_e \to \nu_{sterile}$ oscillations
(and/or maximal $\nu_\mu \to \nu_{sterile}$ oscillations) are
consistent with standard big-bang nucleosynthesis (BBN) for a
large range of parameters \cite{lepasym}.}. This maximum
oscillation solution to the solar neutrino problem can explain the
50 $\%$ flux reduction for a large parameter range:
\begin{eqnarray} 3 \times 10^{-10} \hbox{ eV}^2 \stackrel{<}{\sim}
\delta m^2 \stackrel{<}{\sim} 10^{-3} \hbox{ eV}^2.\end{eqnarray}
The upper bound arises from the lack of $\bar{\nu}_e$
disappearance in the CHOOZ and Palo Verde experiments
\cite{CHOOZ}, while the lower bound comes from a lack of any
distortion in the measured Super-Kamiokande recoil energy spectrum
\cite{skrec,skrec2} , which should make its appearance for $\delta
m^2 \stackrel{<}{\sim} 3 \times 10^{-10}\ \hbox{eV}^2$
(traditional `just so' region). Note that, the maximum oscillation
solution was the only oscillation solution to predict the
approximate energy independent spectrum obtained by
Super-Kamiokande. (SMA MSW, LMA MSW and `Just So' all predicted
some distortion that should have been seen.) The Super-Kamiokande
collaboration has also searched for a day-night effect. However no
evidence for any difference in the day and night time event rates
were found with a $3\sigma$ upper limit of \cite{skrec}
\begin{equation}
A_{n-d} < 0.055
\end{equation}
where $A_{n-d} \equiv (N - D)/(N + D)$ (N = night time events and
D = day time events). This limit allows a slice of parameter space
to be excluded (using the numerical results of Ref.\cite{cfv}):
\begin{eqnarray}
2\times 10^{-7}  \hbox{ eV}^2 &\stackrel{<}{\sim}& |\delta
m^2_{solar}| \stackrel{<}{\sim} 10^{-5} \hbox{ eV}^2 \ \ \ \ \
\hbox{  (sterile) } \nonumber \\ 4\times 10^{-7}  \hbox{ eV}^2 &
\stackrel{<}{\sim}& |\delta m^2_{solar}| \stackrel{<}{\sim}
2\times 10^{-5} \hbox{ eV}^2 \ \hbox{(active) }.
\end{eqnarray}
Thus for both the active and sterile maximal oscillation solutions
the allowed $\delta m^2$ range breaks up into a high $\delta m^2$
region and a low $\delta m^2$ region:
\begin{eqnarray}
2y \times 10^{-5} &\stackrel{<}{\sim}& \delta m^2/\hbox {eV}^2
\stackrel{<}{\sim}  10^{-3} \ \ \ \ \ \ \ \  \hbox{(high} \ \delta
m^2 \ \hbox{region)} \nonumber \\ 3\times 10^{-10}
&\stackrel{<}{\sim}& \delta m^2/\hbox {eV}^2 \stackrel{<}{\sim} 4y
\times 10^{-7} \ \hbox{(low} \ \delta m^2 \ \hbox{region)},
\end{eqnarray}
where $y = 0.5$ for sterile case and $y=1$ for the active case.
These oscillation solutions will be tested in the near future 
by SNO, Borexino and KamLAND experiments.  In fact the SNO
experiment has recently announced their first results\cite{snor}
which is a measurement of the charged current event rate.
This result, when combined with the elastic scattering rate obtained
at super-Kamiokande disfavours the $\nu_e \to \nu_{sterile}$
oscillation solution
at about the $3$ sigma level\cite{snor}. However both measurements
are dominated by systematics which suggests that
this result is not yet convincing but is nevertheless
an interesting hint.
Things should become much clearer when SNO measures the neutral current
event rate
which should allow $\nu_e \to \nu_\alpha$ to be distinguished
from $\nu_e \to \nu_{sterile}$ at more than 7 sigma.
Meanwhile, Borexino
\cite{Borexino} can test the low $\delta m^2$ region by searching
for a day-night effect and also seasonal effects\cite{sg} while
KamLAND \cite{KamLand} can test the high $\delta m^2$ region by
searching for $\nu_e$ disappearance. Finally part of the high
$\delta m^2$ region, $10^{-4} \stackrel{<}{\sim} \delta
m^2/\hbox{eV}^2 \stackrel{<}{\sim} 10^{-3}$ impacts on the
atmospheric electron-like events and is currently
disfavoured\cite{bunn}.

\subsection{LSND data}

There is strong and interesting evidence for $\bar \nu_\mu
\rightarrow \bar \nu_e$ oscillations coming from the LSND
experiment \cite{LSND} which suggests the parameter region
\begin{eqnarray}0.2 \stackrel{<}{\sim} \delta m_{L\!S\!N\!D}^2/
\hbox{eV}^2 \stackrel{<}{\sim} 3
\end{eqnarray}
with a rather small mixing angles $\sin^2 2\theta \sim 3 \times
10^{-2} - 10^{-3}$. This result will be checked by BooNE
\cite{BooNE} soon. If the LSND signal is verified then one must
invoke additional (sterile) neutrino(s) to simultaneously explain
all the solar, atmospheric and LSND data due to the large $\delta
m^2_{L\!S\!N\!D} \sim$ eV$^2$ gap. Even if LSND is not verified,
effectively sterile neutrino may still be responsible for the
solar and/or atmospheric neutrino anomalies. The origin of the
atmospheric and solar neutrino anomalies is something that only
careful experimental studies can establish.

\section{The model}
In this paper we shall study the physics of the neutrino masses in
the alternative 422 model \cite{422.1,422.2}. Before doing so it
is instructive to revise the essentials of this model (with some
refinement). The gauge symmetry of the alternative 422 model is
\begin{equation}
SU(4) \otimes SU(2)_L \otimes SU(2)_R. \label{1}
\end{equation}
Under this gauge symmetry the fermions of each generation
transform in the anomaly free representations:
\begin{equation}
Q_L \sim (4,2,1),\  Q_R \sim (4, 1, 2), \ f_L \sim (1,2,2).
\label{2}
\end{equation}
The minimal choice of scalar multiplets which can both break the
gauge symmetry correctly and give all of the charged fermions mass
is
\begin{equation}
\chi_L \sim (4, 2, 1), \ \chi_R \sim (4, 1, 2),\ \phi \sim
(1,2,2). \label{3}
\end{equation}
Observe that the required scalar multiplets have the same gauge
representation as those of the fermions which gives some degree of
elegance to the scalar sector (although there are three
generations of fermions and only one generation of scalars). These
scalars couple to the fermions as follows:
\begin{equation}
{\cal L} = \lambda_1 \hbox {Tr} \bigg[ \overline{Q_L} (f_L)^c
\tau_2 \chi_R\bigg] + \lambda_2 \hbox {Tr} \bigg[\overline {Q_R}
f^T_L \tau_2 \chi_L \bigg]+ \lambda_3 \hbox {Tr} \bigg[\overline
{Q_L} \phi \tau_2 Q_R\bigg]  + \lambda_4 \hbox {Tr}
\bigg[\overline {Q_L} \phi^c \tau_2 Q_R\bigg] + \hbox{H.c.},
\label{4}
\end{equation}
where the generation index has been suppressed and $\phi^c =
\tau_2 \phi^* \tau_2$.  Under the $SU(3)_c \otimes U(1)_T$
subgroup of $SU(4)$, the $4$ representation has the branching rule
$4 = 3(1/3) + 1(-1)$.  We will assume that the $T=-1, I_{3R} = 1/2
\ (I_{3L}=1/2)$ components of $\chi_{R} (\chi_L)$ gain non-zero
Vacuum Expectation Values (VEVs) as well as the $I_{3L} = -I_{3R}
= -1/2$ and $I_{3L} = - I_{3R} = 1/2$ components of the $\phi$. We
denote these VEVs by $w_{R,L}, u_{1,2}$ respectively. In other
words,
\begin{eqnarray}
\langle \chi_R (T = -1, I_{3R} = 1/2) \rangle = w_R, \ \langle
\chi_L (T = -1, I_{3L} = 1/2) \rangle = w_L, \nonumber \\ \langle
\phi (I_{3L} = -I_{3R} = -1/2)\rangle = u_1,\ \langle \phi (I_{3L}
= -I_{3R} = 1/2)\rangle = u_2.
\end{eqnarray}
We will assume that the VEVs satisfy $w_R > u_{1,2}, w_L$ so that
the symmetry is broken as follows:
\begin{eqnarray}
&SU(4)\otimes  SU(2)_L \otimes SU(2)_R&
 \nonumber \\
&\downarrow \langle \chi_R \rangle& \nonumber \\ &SU(3)_c \otimes
SU(2)_L \otimes U(1)_Y & \nonumber \\ &\downarrow \langle \phi
\rangle, \langle \chi_L \rangle \nonumber \\ &SU(3)_c \otimes
U(1)_Q&
\end{eqnarray}
where $Y = T +2I_{3R}$ is the linear combination of $T$ and
$I_{3R}$ which annihilates $\langle \chi_R \rangle$ (i.e.
$Y\langle \chi_R \rangle = 0$) and $Q = I_{3L} + {Y \over 2}$ is
the generator of the unbroken electromagnetic gauge symmetry.
Observe that in the limit where $w_R \gg w_L, u_1, u_2$, the model
reduces to the standard model. The VEV $w_R$ breaks the gauge
symmetry to the standard model subgroup.

To facilitate easy reference, we will use $\alpha =\pm {1 \over 2}
$, $\beta =\pm {1 \over 2}$ to index the $SU(2)_L$ and $SU(2)_R$
component respectively, whereas $\gamma \equiv \{\gamma', 4\}$ is
used to index the $SU(4)$ components, where $\gamma' \equiv
(y,g,b)$ is the usual colour index for $SU(3)_c$, and $\gamma = 4$
the forth colour. With this index scheme the fermion multiplets
are written as (with the generation index suppressed)
\begin{eqnarray}
Q_L^{\alpha, \gamma} = \left(\begin{array}{cccc} \tilde U &
{\tilde E}^0  \\ D & \tilde E^- \end{array}\right)_L, \ 
Q_R^{\beta,\gamma} = \left(\begin{array}{cc} 
\tilde U & \tilde \nu \\ \tilde D
& \em l \\
\end{array}
\right)_R, \ f_L^{\alpha, \beta} = \left(\begin{array}{cc} (\tilde
E^-_R)^c & \tilde \nu_L \\ (\tilde E^0_R)^c & \em l_L
\end{array}
\right). \label{multi}
\end{eqnarray}
The rationale to label some of the fermion fields in the
multiplets with a tilde will be addressed shortly in the next
paragraph. In the above matrices the first row of $Q_L$ and $f_L \
(Q_R)$ is the $I_{3L} \ (I_{3R}) = \alpha \ (\beta)= 1/2$
component while the second row is the $I_{3L} \ (I_{3R}) = \alpha
\ (\beta) = -1/2$ component. The two columns of $Q_L, Q_R$ are the
$\gamma = \gamma'$ and $\gamma = 4$ components of $SU(4)$, and the
columns of $f_L$ are the $I_{3R} = \beta = \pm 1/2$ components.
Each field in the multiplets Eq. (\ref{multi}) represents $3\times
1$ column vector of three generations.

As in the case of the Standard Model, the flavour states of the
fermion fields are, in general, not aligned with the corresponding
mass eigenstates. This shall necessitate the introduction of some
theoretically arbitrary CKM-type unitary matrices into the theory.
Without loss of generality we can choose a basis such that $D_L$
(the left-handed down-type quarks) in $Q_L$, $\em l_R$ (the
right-handed charged leptons fields) in $Q_R$ and $\em l_L$
(left-handed charged lepton fields) in $f_L$ are (almost
\footnote{ Strictly, ${\em l}_L, {\em l}_R$ are approximately but
not exactly mass eigenstates because of the mass mixing between
${\em l}_R$ and $\tilde E^-_L$ [see the forthcoming Eq.
(\ref{ybyz})]. This means that the true charged lepton mass
eigenstate fields have the form ${\em l}^{true}_L = {\em l}_L +
{\cal O}({M_d M_{\em l} \over M^2_{E}})E_L, {\em l}^{{true}}_R =
{\em l}_R + {\cal O}({M_d \over M_{E}})E_R$.}) mass eigenstate
fields that couple with their respective diagonal mass matrix. The
rest of the fields $\tilde \psi$ (fields with a tilde) are flavour
states.

It is instructive to list and label the scalar fields index
explicitly as follows:
\begin{eqnarray}
\chi_L^{\alpha, \gamma} = \left(\begin{array}{cccc}
\chi_L^{\gamma', {1 \over 2}} & \chi_L^{4, {1 \over 2}} \\
\chi_L^{\gamma',-{1 \over 2}} &  \chi_L^{4, -{1 \over 2}}
\end{array}\right),\ \chi_R^{\beta, \gamma} =
\left(\begin{array}{ccc} \chi_R^{\gamma', {1 \over 2}} &
\chi_R^{4, {1 \over 2}} \\ \chi_R^{\gamma',-{1 \over 2}} &
\chi_R^{4, -{1 \over 2}}
\end{array}\right),\
\phi^{\alpha, \beta} = \left(\begin{array}{ccc} \phi^{{1 \over 2},
{1 \over 2}} & \phi^{{1 \over 2}, -{1 \over 2}}\\ \phi^{-{1 \over
2}, {1 \over 2}} & \phi^{-{1 \over 2}, -{1 \over 2}}
\end{array}\right).
\label{scalar}
\end{eqnarray}
The electric charges of the components of $\chi_{L,R}$, $\phi$ can
be read off from Eq. (\ref{multi}) because the scalars and
fermions have the same gauge representation.

The mass matrices (after spontaneous symmetry breaking) of the
fermions are derived from the Yukawa Lagrangian of Eq. (\ref{4})
as follows:
\begin{equation}
{\cal L} \ ({\hbox{after SSB}})
 = {\cal L}^E + {\cal L}^{\em l}
+ {\cal L}^u + {\cal L}^{d} + \hbox{H.c.}, \label{LM}
\end{equation}
where
\begin{eqnarray}
-i{\cal L}^E = \overline {{\tilde E}^-_L} {\tilde M}_E \tilde
{E}^-_R - \overline {{\tilde E}^0_L}{\tilde M}_E \tilde {E}^0_R ,
\ -i {\cal L}^{\em l} =  \overline {({\tilde E}_R^0)^c} M_{\em
l}{\tilde \nu}_R  + \bar {\em l}_L M_{\em l}{\em l}_R, \nonumber
\\ -i {\cal L}^u  = \overline {{\tilde U}_L}{\tilde
M}_u{\tilde U}_R + \overline {\tilde{E}_L^0}{\tilde M}_u\tilde
{\nu}_R, \ -i {\cal L}^d = - \overline {D_L}\tilde{M}_d{\tilde
D}_R - \overline {\tilde E^-_L} \tilde{M}_d{\em l}_R, \label{ybyz}
\end{eqnarray}
and
\begin{equation}
\tilde M_E = w_R\lambda_1, \: \tilde{M}_u=(\lambda_3 u_2 -
\lambda_4 u_1), \: \tilde{M}_d=(\lambda_4 u_2 - \lambda_3 u_1),
\end{equation} are $3\times3$ generally
non-diagonal mass matrices for the exotic $E$ leptons, up-type
quarks $U$ and down-type quarks $D$ respectively.
\begin{equation}
M_{\em l} = w_L\lambda_2 = \left( \begin{array}{cccc} m_{\em_e}&
0&0 \\ 0&m_{\em_\mu}&0\\ 0&0&m_{\em_\tau}\end{array}\right)
\end{equation}
is the diagonal mass matrix for the charged leptons. The mass
matrices ${\tilde M}_E$, ${\tilde M}_u$, ${\tilde M}_d$ can be
diagonalised in the usual way (biunitary diagonalisation) to
obtain the diagonal mass matrices, e.g. $M_E = U^\dagger \tilde
M_E V $ etc. so that the flavour states $\tilde \psi$ are related
to their mass eigenstates $\psi$ via unitary matrices:
\begin{eqnarray}
E_R &=& V^\dagger {\tilde E}_R, \;\;\ E_L = U^\dagger {\tilde
E}_L, \nonumber \\ U_R &=& Y_R^\dagger{\tilde U}_R,\;\;\ U_L =
Y_L^\dagger{\tilde U}_L, \nonumber \\ D_R &=& K'{\tilde D}_R,
\label{basis}
\end{eqnarray}
and $\tilde D_L = D_L, \tilde {\em l}_L = {\em l}_L, \tilde {\em
l}_R = {\em l}_R$ with our choice of basis. Note that
approximately the same unitary matrix $U$ ($V$) relates both
$E_L^0$, $E_L^-$ ($E_R^0$, $E_R^-$) to their weak eigenstates
$\tilde E_L^0$, $\tilde E_L^-$ ($\tilde E_R^0$, $\tilde E_R^-$)
because the mass matrices are approximately $SU(2)_L$ ($SU(2)_R$)
invariant.

The gauge fields in the 15 representation of $SU(4)$ has the
branching rule $15 = 8(0)+ 3(-{4\over3})+ \bar 3 ({4\over3}) +
1(0)$ under $SU(3)_L \otimes U(1)_T$. We identify the colour octet
8(0) as the gluons of the usual $SU(3)_c$ colour group,
$3(-{4\over3}),\bar 3({4\over3})$ colour triplet lepto-quark gauge
bosons $W', W'^*$ that couple the $\gamma = \gamma'$ components to
$\gamma = 4$ component in $\chi_{L,R}$, $Q_{L,R}$. Note that there
is also a neutral gauge boson $B'_\mu$ corresponds to the singlet
1(0). The matrix $Y_L^{\dagger} \equiv K_L$ is the usual CKM
matrix (as in the Standard Model), whereas $Y_R^{\dagger}
K'^{\dagger} \equiv K_R$ is the analogue of the CKM matrix for the
right-handed charged quarks in $SU(2)_R$ sector. The matrix $K'$
is the analogue of the CKM-type matrix in the $SU(4)$ sector
pertaining to lepto-quark interactions mediated by $W', W'^*$. It
was shown in Ref. \cite{422.1,422.2} that the main experimental
constraints on this 422 model come from rare $B$ and $K$ decays
such as $K^0 \to \mu^\pm e^\mp, B^0 \to \mu^\pm \tau^\mp$ etc.
depending on the form of $K'$. Remarkably, the symmetry breaking
scale could be as low as a TeV without being in conflict with any
experimental measurements. Indeed we will assume that the symmetry
breaking scale is in the interesting low range: 
\begin{equation}\label{Mrange}
 0.5 (1.0)\hbox{ TeV } \stackrel{<}{\sim} M_{W_R}
  (M_{W'})\stackrel{<}{\sim} 10\hbox{ TeV}, \ \nonumber \\
  45 \hbox{ GeV } \stackrel{<}{\sim} M_{E} \stackrel{<}{\sim}
  10\hbox{ TeV},
\end{equation}
which is well motivated because it avoids the gauge hierarchy
problem and it also allows the model to be testable at existing
and future colliders (such as LHC).
(Note that the lower limit on the mass of the $E$ lepton arises from LEP
measurements of the $Z^0$ width).

Apart from laboratory experimental bounds, there are also
astrophysical ones on $M_{W_R}$. In particular 
Ref. \cite{Raffelt,Barbieri,Lang} have
argued that based on `energy-loss' argument, Supernova 1987A
excludes a range of values for $M_{W_R}$ and the $W_L - W_R$
mixing parameter, $\zeta$ \footnote {In our case here, $\zeta
\simeq {\mu^2 \over M^2_{W_R}} \stackrel{<}{\sim} 4 \times 10^{-5} -
1 \times 10^{-4}$, where $\mu^2 = g_L g_R u_1 u_2$, see forth
coming Eq. (\ref{whcl}).} (for $m_{\nu_R} \stackrel{<}{\sim} 10$
MeV):
\begin{eqnarray}
\zeta &<& 10^{-5}, \nonumber \\ (0.3 - 0.5) &\stackrel{<}{\sim}
&{M_{W_R}\over \hbox{ TeV}} \stackrel{<}{\sim} 22 - 40 \,\,\,\,
\hbox{ in the limit }{\zeta \rightarrow 0}, \label{1987}
\end{eqnarray}
which seems to marginally rule out the range of $M_{W_R}$ as
assumed in Eq. (\ref{Mrange}). 
However, while there will undoubtly be effects for
supernova coming from the additional gauge bosons in our
model, we should nonetheless keep in mind
that the modeling of core collapse of supernova is generally
plagued by theoretical, observational as well numerical
uncertainties. For example, Berezinsky\cite{Berezinsky} has 
recently emphasised that the reasonably successful description of
SN1987A is somewhat surprising given that it was assumed that
the presupernova protostar was a non-rotating red super giant,
while it appears that it was actually a rotating blue super giant.
Also, Turner \cite{Turner} pointed
out that there exists uncertainty in the theoretical model for the
hot core of a core collapsed supernova, which itself depends
critically upon the equation of state at supernuclear densities
(which is a state-of-art problem in nuclear physics). The
criterion of neutrino luminosity from SN 1987A, $Q_a
\stackrel{<}{\sim} 10^{53}$ erg s$^{-1}$, a key ingredient in
obtaining the SN 1987A bound of Eq. (\ref{1987}), according to
Ref. \cite{Turner} is also subjected to question. Raffelt and
Seckel \cite{Raffelt} pointed out that SN 1987A bound on the
right-handed and other light exotic particle interactions could be
uncertain up to as much as 2 order of magnitude, in which case
could render the apparent `conflict' of the range assumed by Eq.
(\ref{Mrange}) and the SN 1987A bound of Eq. (\ref{1987}) be
alleviated. While keeping in mind the possible astrophysical 
implications of a
low symmetry breaking scale $M_{W_R} \sim 1$ TeV, we now 
continue with our
exploration of the possible phenomenology of the alternative 422
model.

The $SU(2)_{L,R}$ charged gauge bosons $W_{L,R}^\pm$ couple to the
fermions via the interaction Lagrangian density
\begin{eqnarray}
i{\cal L}^{gauge} = & &{g_L \over \sqrt 2} \lbrack \overline
{{U}_L} \ {/ \!\!\!\! {W}^+_L}K_L  D_L + \overline {{\tilde
\nu}_L} \ {/ \!\!\!\! {W}^+_L} {\em l}_L + \overline {E^0_L} \ {/
\!\!\!\! {W}^+_L} E^-_L  + \overline {(E^-_R)^c}\ {/ \!\!\!\!
{W}^+_L} (E^0_R)^c + \hbox{ H.c.}\rbrack \nonumber \\ &+ &{g_R
\over \sqrt 2} \lbrack \overline{{U}_R} \ {/ \!\!\!\! {W}^+_R} K_R
D_R + \overline{{\tilde \nu}_R} \ {/ \!\!\!\! {W}^+_R} {\em l}_R
+\overline {(E^0_R)^c}V^{\dagger}\ {/ \!\!\!\! {W}^+_R} {\em l}_L
+ \overline {(E^-_R)^c} \ V^{\dagger}\ {/ \!\!\!\! {W}^+_R}
{\tilde \nu}_L  \nonumber \\ &+& \hbox{ H.c.}\rbrack. \label{LG}
\end{eqnarray}
The charged gauge interactions are of our interest because they
will give rise to radiative Dirac mass terms to the neutrinos in
this model as we will now discuss.

\section{Neutrino mass}
With the model as defined in Section III, the ordinary neutrinos
$\tilde \nu_L$ are massless at tree level because the $\tilde
\nu_L$ states do not couple to any VEV \lbrack see Eq.
(\ref{LM})\rbrack \footnote{To generate tree-level neutrino mass
we need to either admit the gauge invariant bare mass term
$m_{bare} \bar f_L(f_L)^c$ into the Lagrangian density in Eq.
(\ref{4}) or add a new Higg $\Delta \sim (4,2,3)$ into the
Lagrangian density via the coupling $\lambda_{\Delta}
\Delta^\dagger \overline{Q_R}f_L + $H.c.. By developing a VEV,
$\Delta$ can generate a Dirac mass term $\overline {\tilde \nu_L}
m_D \tilde \nu_R$. We argue that, since the scale of $m_{bare}$ is
completely independent of the weak scale, the assumption that
$m_{bare} \ll M_{weak}$ is surely an interesting possibility. In
view of this plausible assumption we have set $m_{bare} = 0$.
Meanwhile, adding an additional scalar multiplet such as $\Delta$
spoils both the simplicity and elegance of the model.}. However,
the neutrino masses are non-zero in the model because there are
1-loop (and higher order) Feynman diagrams which contribute to
their masses. In other words the masses of the neutrinos arise
radiatively in the model. In particular, as we will see later, a
Dirac mass $m_D$ and a $\tilde \nu_L-(E_L)^c$ mass mixing term
$m_{\nu E}$ will be generated as mass corrections at 1-loop level,
meanwhile the $\tilde \nu_R$ states gain Majorana masses at
tree-level by mixing with $E$ leptons. We will elaborate these
mechanisms in more detail in the following subsections. In this
paper we will be working exclusively in the t'Hooft-Feynman gauge.

\subsection{Tree level Majorana mass matrix $M_R$}
At tree level, mixing between $\tilde \nu_R$ with $E_{L,R}^0$
generates right-handed neutrino Majorana mass $M_R$. In the mass
eigenstate basis (for the $E^0_{L,R}$) defined in Eq.
(\ref{basis}), the tree-level Lagrangian density of Eq. (\ref{LM})
becomes
\begin{eqnarray}
-i{\cal L} &=& {1 \over 2} \left( \begin{array}{cccc} \overline
{{\tilde \nu}_L} & \overline {({\tilde \nu}_R)^c} &\overline
{E^0_L}&\overline {(E^0_R)^c}\end{array} \right){\bf M} \left(
\begin{array}{cccc} ({{\tilde \nu}_L})^c \\ {\tilde
\nu}_R\\(E^0_L)^c \\(E^0_R)
\end{array} \right) \nonumber \\
& & - \left( \begin{array}{cccc} \overline {D_L} &\overline
{E^-_L}\end{array} \right) \left(\begin{array}{cccc} M_d&0\\ 0&
U^{\dagger} M_d K' \end{array} \right) \left( \begin{array}{cccc}
D_R\\{\em l}_R \end{array} \right) \nonumber \\ & & + \overline
{E^-_L}M_E{E^-_R} + \bar {\em l}_L M_{\em l}{\em l}_R + \overline
{U_L} M_u U_R + \mbox{H.c.}, \label{lm2}
\end{eqnarray}
where
\begin{eqnarray}
{\bf M} &=& \left( \begin{array}{cccc} 0&0&0&0\\ 0&0 & Y_R M_u
Y_L^\dagger U & M_{\em l}V\\ 0&(Y_R M_u Y_L^\dagger
U)^\dagger&0&-M_E\\ 0& (M_{\em l}V)^\dagger&-M_E&0
\end{array}\right) \;\;\; (12 \times 12 \hbox{ matrix}). \label{yyt2}
\end{eqnarray}
$M_d$, $M_u$ are the $3 \times 3$ diagonal mass matrices for the
down-type quarks and up-type quarks respectively, whereas $M_E$ is
the $3 \times 3$ diagonal mass matrix for $E$ leptons after
biunitary-diagonalising $\tilde M_E = w_L\lambda_1$:
\begin{equation}
\tilde M_E = U M_E V^\dagger \equiv U \left( \begin{array}{cccc}
M_{E_1}&0&0\\ 0&M_{E_2}&0 \\ 0&0&M_{E_3}
\end{array}\right) V^\dagger.
\label{lambda1}
\end{equation}
The matrices $U, V$ describe the relation between the weak and
mass eigenstates of the $E$ leptons and can be determined from
$\lambda_1$ [see Eq. (\ref{basis})].

One can block diagonalise the $9\times 9$ matrix in the lower
right sector of ${\bf M}$ by a similarity transformation \cite{st}
using the approximately orthogonal matrix
\begin{eqnarray}
U_\rho &=& \left(
\begin{array}{cccc}{\bf I}_3 & \rho_p \\ -\rho_p^T & {\bf I}_6 \end{array}
\right) \ \ (9 \times 9 \hbox{ matrix}) \nonumber
\end{eqnarray} where
\begin{eqnarray}
\rho_p &=& \left(\begin{array}{cccc} Y_R M_u Y_L^\dagger U &
M_{\em l}V \end{array} \right) \left(\begin{array}{cccc} 0 &
-M^{-1}_E \\ -M^{-1}_E & 0 \end{array} \right). \ \ (3 \times 6
\hbox{ matrix})
\end{eqnarray}
Block diagonalisation casts ${\bf M}$ into the form
\begin{eqnarray}
{\bf M} \rightarrow \pmatrix{ \begin{array}{cccc} 0 & 0 \\ 0 & M_R
\end{array} \  \begin{array}{cccc}  0 \, \, & 0 \, \,
\\ 0 \, \, & 0 \, \,
\end{array} \cr \begin{array}{cccc}0 & \ 0 \
\\ 0  &  \ 0  \
\end{array} \ \matrix{-\bf M_E'}},
\end{eqnarray}
where each `0' is a $3 \times 3$ matrix of zeros, and
\begin{eqnarray}
M_R &\simeq& \left(M_{\em l}V M_E^{-1} U^\dagger Y_L M_u
Y_R^\dagger \right) + \left(M_{\em l}V M_E^{-1} U^\dagger Y_L M_u
Y_R^\dagger\right)^\dagger, \ \ (3 \times 3 \hbox { matrix})
\nonumber
\\ \bf {M_E'} & \simeq & \pmatrix {0&M_E\cr M_E&0}.
\ \ (6 \times 6 \hbox { matrix}) \label{gMR}
\end{eqnarray}
In the limit $M_E \gg M_{\em l}, M_u$, the states $\tilde \nu_R$
are decoupled from the $E$ leptons. In the special case of
decoupled generations (e.g. $ Y_L=Y_R = U = V = \bf I$), ${\bf M}$
reduces to
\begin{equation}
{\bf M} = \left(\begin{array}{cccc} 0&0&0&0\\ 0&0&m_{q_u}&m_{\em
l}\\ 0&m_{q_u}&0&-M_{E_i}\\ 0&m_{\em l}&-M_{E_i}&0
\end{array}\right),
\label{yyt0}
\end{equation}
where $i$, $q_u$ and $\em l$ index the generations. In this case,
Eq. (\ref{gMR}) reproduces the result
\begin{equation}
M_R = {2m_{q_u} m_{\em l} \over M_{E_i}} \label{MR}
\end{equation}
as obtained in Ref. \cite{422.2}.

\subsection{Radiative correction to $\bf M$ due to gauge interactions}

At the 1-loop level the gauge interactions from the charged
$SU(2)_L$ gauge bosons $W_L^{\pm}$ and $SU(2)_R$ gauge bosons
$W_R^{\pm}$ give rise to $\overline {\tilde \nu_L}(\tilde\nu_L)^c$
Majorana mass $m_M$, $\overline {\tilde \nu_L} \tilde \nu_R$ Dirac
mass $m_D $ and $\overline {\tilde \nu_L}(E^0_L)^c$ mass mixing
term $m_{\nu E}$. The Dirac mass $m_D$  arises from the gauge
interactions\footnote{The (unphysical) Goldstone boson
contributions will be evaluated together with the (physical)
scalar contributions in Part C of this section.}
\begin{equation}
{g_L \over \sqrt 2} \overline {\tilde \nu_L} W^+_{\mu L}{\em l}_L
+ {g_R \over \sqrt 2} \overline {\tilde \nu}_R W^+_{\mu R}{\em
l}_R + \hbox{ H.c.}
\end{equation}
which leads to the Feynman diagram: \vskip 0.8 cm
\centerline{\epsfig{file=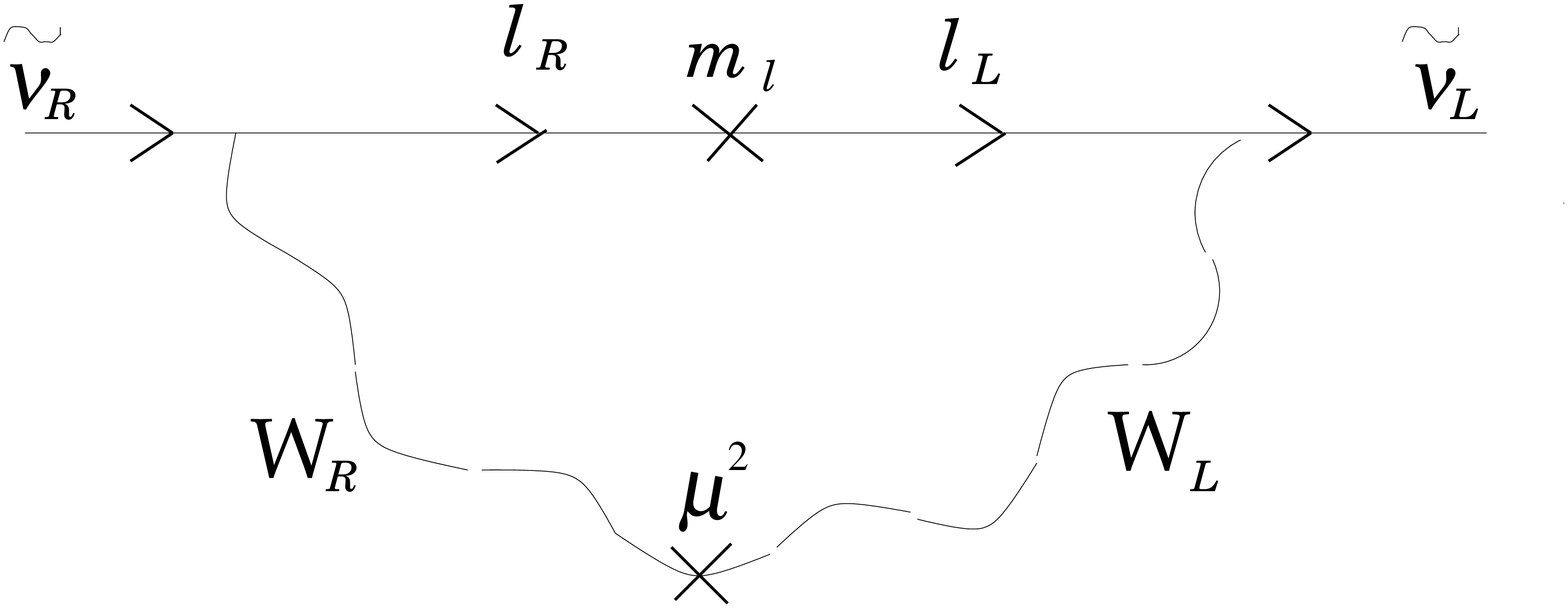,width=9cm}}
\vskip 0.5cm
\centerline{ {\bf
Figure 1}.$\;$ Dirac mass generated by gauge interactions leading
to the mass term $\overline{\tilde \nu_L}\tilde \nu_R$.} \vskip
0.5 cm \noindent Note that since these interactions do not mediate
cross generational mixing among the neutrinos, the Dirac mass
matrix $m_D$ is strictly diagonal. As calculated in Ref.
\cite{422.2}, for each generation,
\begin{eqnarray}
m_D = m_{\em l} {g_Rg_L\over8\pi^2}{\mu^2\over M^2_{W_R}} {\ln
\left( {M^2_{W_R} \over M^2_{W_L}} \right)} \equiv m_{\em l}k_g,
\label{md}
\end{eqnarray}
where $\mu^2 \equiv g_Lg_Ru_1u_2$ is the $W_L - W_R$ mixing mass.
Note, that the lower limit of $\mu^2$ is simply zero since $\mu^2$ 
vanishes in either
limits of $u_1 \rightarrow 0$ or $u_2 \rightarrow 0$ and the
theory remains phenomenologically consistent in this limit.
It was also shown in the Ref.\cite{422.2} that 
\begin{equation}
{\mu^2 \over M^2_{W_R}} \stackrel{<}{\sim} {1 \over 2\sqrt3 }
{M^2_{W_L} \over M^2_{W_R}}{m_b \over m_t}, \label{whcl}
\end{equation}
which is not a strict upper limit, but rather an approximate
condition to avoid fine tuning \footnote {As shown in Ref.
\cite{422.2}, the limit of Eq. (\ref{whcl}) comes from ${u_1u_2
\over u_1^2 + u_2^2} \stackrel{<}{\sim} {m_b \over m_t}$. To avoid
the equality $m_b = m_t$ we will need the scale of $u_1, u_2$ be
separated by a hierarchy, else we will need to fine-tune the
Yukawa coupling constants $\lambda_{3,4}.$}. In view of this rough
upper limit, it is convenient to write $m_D$ in terms of a
parameter $ 0<\eta<1$ defined such that
\begin{equation}
m_D = m_{\em l}k_g = m_{\em l} \eta S \label {mddcpl}
\end{equation}
where
\begin{eqnarray}
S &=& S(M_{W_R}) = {g_Rg_L\over8\pi^2}{\ln \left( {M^2_{W_R} \over
M^2_{W_L}} \right)} {1 \over 2\sqrt3 } {M^2_{W_L} \over
M^2_{W_R}}{m_b \over m_t} \nonumber \\ & \sim & 10^{-7} \left(
{\hbox{TeV} \over M_{W_R}}\right)^2.
\end{eqnarray}
If we take the reasonable range of $0.5 \hbox{ TeV} <M_{W_R}
\stackrel{<}{\sim}\hbox{10 TeV}$, then $S$ typically spans a range
of
\begin{equation} 10^{-6} < S \stackrel{<}{\sim}
10^{-9} \label{Srange}.\end{equation} Thus, the gauge loop
contribution to $m_D$ is proportional to $m_{\em l}$ and is
naturally light because of its radiative origin.

The mass mixing term $m_{\nu E}$ arises at 1-loop level via the
gauge interactions
\begin{equation}
{g_L \over \sqrt 2} \overline {E^0_L} W^+_{\mu L}E^-_L + {g_R
\over \sqrt 2} \overline {(E^-_R)^c}V^\dagger W^+_{\mu R}{\tilde
\nu}_L + \hbox{H.c.} \label{des}
\end{equation}
leading to the Feynman diagram: \vskip 0.8 cm
\centerline{\epsfig{file=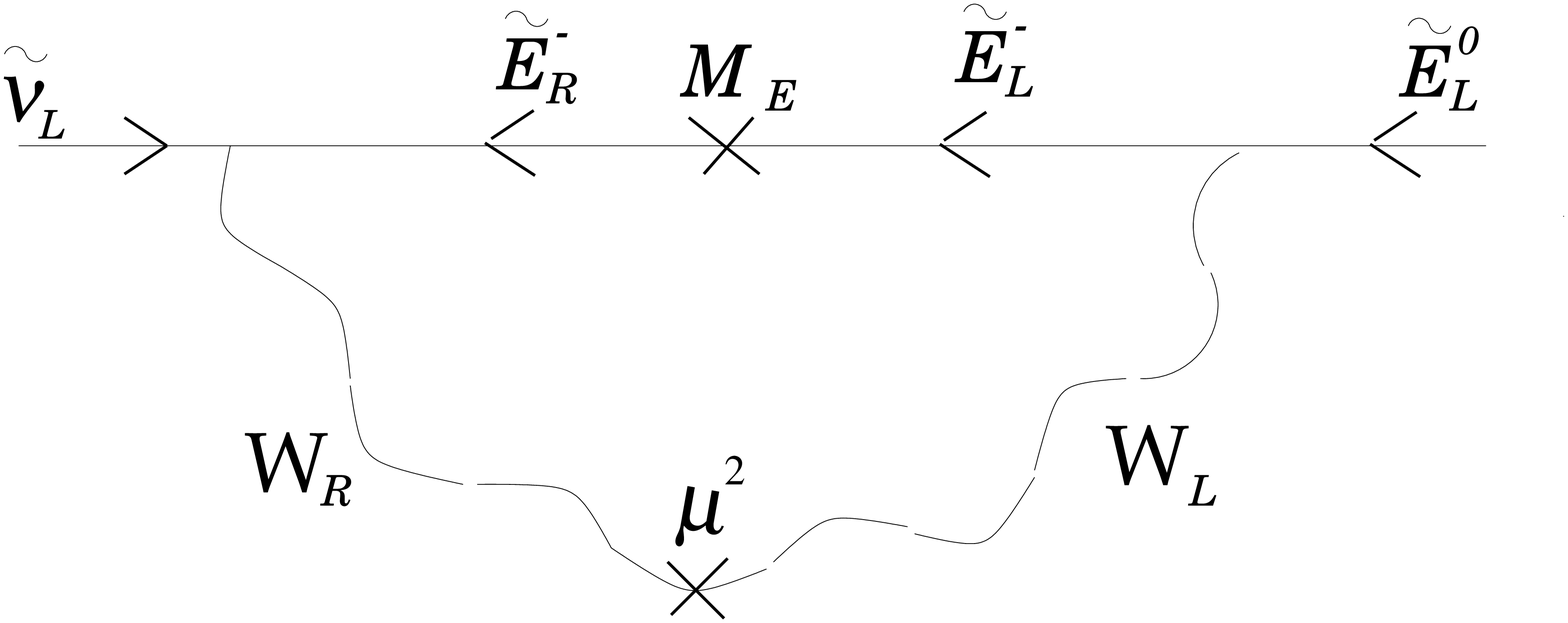,width=9cm}}
\vskip 0.5cm
\noindent {\bf
Figure 2}. $\,$ $\overline{\tilde \nu_L}(\tilde E^0_L)^c$ neutrino
mixing term generated by gauge interactions leading to the mass
term $m_{\nu E}$. \vskip 0.5 cm \noindent Note that in contrast to
$m_D$, the involvement of the matrix $V$ in the interactions Eq.
(\ref{des}) may mediate cross generational mixing. In the special
case of decoupled generations, $m_{\nu E}$ has been calculated
earlier by Ref. \cite {422.2} as
\begin{eqnarray}
m_{\nu E} &=& M_E {g_R g_L \over 8 \pi^2} \left( {\mu^2 \over
M_{W_R}^2} \right) \biggl[ \ln \left( {M_{W_R}^2 \over M_{W_L}^2}
\right) + {{M_E^2 \ln \left( M_{W_L}^2 \over M_E^2 \right)} \over
M_{E}^2 - M_{W_L}^2 } - {{M_E^2 \ln \left( M_{W_R}^2 \over M_E^2
\right)} \over M_E^2-M_{W_R}^2} \biggr] \nonumber \\ &\sim& \eta
M_E S. \label{mdnu}
\end{eqnarray}
In addition to $m_D$ and $m_{\nu E}$, this model also generates a
$\overline{\tilde \nu_L} (\tilde \nu_L)^c$ Majorana mass term
$m_M$ at 1-loop level by charged gauge interactions. In the case
of decoupled generations,
\begin{equation}
m_M = {m_{\em l} m_{q_d} M_E}{g_R g_L \over 8 \pi^2} \left( {\mu^2
\over M_{W_R}^2} \right) \biggl[ {\ln \left({M_{W_R}^2 \over
M_{E}^2} \right)\over M_{W_R}^2-M_{E}^2} -{\ln \left({M_{W_L}^2
\over M_{E}^2}\right) \over M_{W_L}^2-M_{E}^2} \biggr] \nonumber
\sim \eta {m_{\em l} m_{q_d} \over M_E} S
\end{equation}
is generically tiny compared to $m_D$ and $m_{\nu E}$.
\subsection{Radiative correction to $\bf M$ due to scalar interactions}
In the previous subsection we have shown that at 1-loop level the
gauge interactions give rise to Dirac mass $m_D$, mass mixing
correction $m_{\nu E}$ and also $m_M$. Nevertheless, in this model
this is not the only way mass corrections could arise. Other than
the gauge mechanism discussed in the previous subsection, the
Higgs sector also contains interactions that could generate Dirac
mass for the neutrinos as a radiative mass correction. The
relevant interactions involve the negatively charged colour
singlet scalar $\chi_{L,R}^{4,-{1 \over 2}}$:
\begin{eqnarray}
-i{\cal L}_{\chi^{4, -{1 \over 2}}} &=& -i {\cal L}_{\chi_L^{4,
-{1 \over 2}}} -i {\cal L}_{\chi_R^{4, -{1 \over 2}}} \nonumber
\\ &=& {\chi_R^{4, -{1 \over 2}} \over w_R} \overline {E^-_L} M_E
V^{\dagger} ({\tilde \nu}_L)^c -{\chi_L^{4, -{1 \over 2}} \over
w_L}\overline {{\tilde \nu}_R} M_{\em l}V(E^-_R)^c \nonumber \\ &
& - {\chi_R^{4, -{1 \over 2}} \over w_R} \overline {E^0_L} M_E
V^{\dagger}({\em l}_L)^c -{\chi_L^{4, -{1 \over 2}} \over w_L}
\overline {l_R} M_{l}{\tilde \nu}_L + \hbox { H. c.}. \label{Lchi}
\end{eqnarray}
The first two terms of Eq. (\ref{Lchi}) will give rise to a Dirac
mass correction $m_D$ with $E^-$ as propagator (see Fig. 3),
whereas the last two terms will give rise to a mass mixing term
$m_{\nu E}$ with charged leptons as propagator. Therefore in
comparison to $m_D$ the mass $m_{\nu E}$ can be safely ignored.
\vskip 0.7 cm \centerline{\epsfig{file=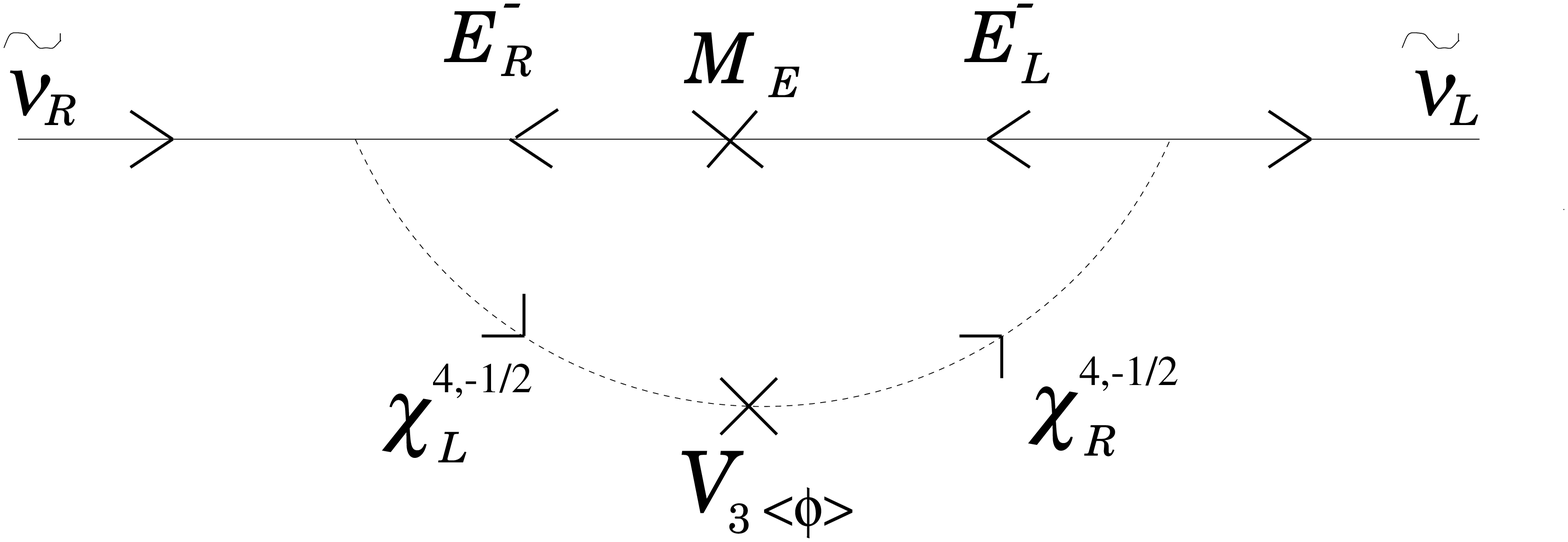,width=9cm}}
\vskip 0.5cm
\noindent {\bf Figure 3}.$\;$ $\overline {\tilde \nu_L}\tilde
\nu_R$ Dirac mass generated by scalar interactions. The cross in
the scalar propagator is the perturbative mass mixing term that
can be shown to vanish in either limit of $u_{1,2} \rightarrow 0$
\vskip 0.5 cm \noindent The mixing between $\chi_L-\chi_R$ is
effected solely by the mixing term of the form
\begin{eqnarray}
 V_3 &=& M \chi_L^{\dagger} \phi \tau_2 \chi_R
+ M'\chi_L^\dagger \phi^c \tau_2 \chi_R + \hbox { H.c.} \nonumber
\\ \Rightarrow V_{3\langle \phi \rangle} &=& m^2_{3\langle \phi
\rangle}\chi_L^{4,-{1 \over 2}} \chi_R^{4,-{1 \over 2}*} + \hbox {
H.c.}; \nonumber \\ m^2_{3\langle \phi \rangle}&\equiv &-(Mu_1 +
M'u_2)
\end{eqnarray}
in the Higgs potential.

The charged gauge bosons $W^{\pm}_{L,R}$ acquire mass by eating
two colourless, charged would-be Goldstone bosons (let's call them
$G^{\pm}_{R}$ and $G^{\pm}_{L}$) which are linear combinations of
$\chi^{4, -{1 \over 2}}_{L,R}$, $\phi^{{1 \over 2},{1 \over 2}}$
and $\phi^{-{1 \over 2},-{1 \over 2}}$, leaving behind two
physical charged Higgs (which we will call $H^{\pm}_{1,2}$). The
fields $\chi_{L,R}^{4, -{1 \over 2}}$ appearing in Fig. 3 are
linear combination of these would-be Goldstone bosons and physical
Higgses. We shall work out the linear combinations of
$G^{\pm}_{L,R}$ and $H^{\pm}_{1,2}$ in $\chi^{4,-{1 \over
2}}_{L,R}$ so that the Dirac mass arising from these
$\chi$-interactions can be calculated.

The Goldstone bosons associated with each spontaneously broken
symmetry are given by
\begin{equation}
G = \Phi^T T^a \lambda_a, \label{G}
\end{equation}
where $\Phi$ are the Higgs scalars of the theory, $T^a$ the
generators of the broken symmetries and $\lambda_a$ the vacua.
Utilising Eq. (\ref{G}), the would-be Goldstone bosons
$G^{\pm}_{L}$ and $G^{\pm}_{R}$ can be identified by taking $T^a$
as the charged $SU(2)_{L,R}$ generators, $\tau_{\pm}$. The two
states which are orthogonal to $G^{\pm}_{R}$ and $G^{\pm}_{L}$
will be related to the physical Higgs $H^{\pm}_{1,2}$ with masses
denoted by $M_{H_1,H_2}$. We will first work in the limit $u_1
\rightarrow 0$ since we expect $u_1 \ll u_2$ (or $u_2 \ll u_1$,
which is analogous to $u_1 \ll u_2$ in the following analysis). In
this limit it is possible to show that (see Appendix) the weak
eigenstates are related to $G^{\pm}_{L,R}, H^{\pm}_{1,2}$ via the
unitary matrix $U_g$ as
\begin{eqnarray}
\left(\begin{array}{cccc} \chi_R^{4,-{1 \over 2}} \\ \chi_L^{4,
-{1 \over 2}} \\ \phi^{{1 \over 2},{1 \over 2}*} \\ \phi^{ -{1
\over 2},-{1 \over 2}} \end{array}\right) = U_g \pmatrix{G_R^-\cr
G_L^{-} \cr H_1^- \cr H_2^-}, \label{lw1}
\end{eqnarray}
where the matrix $U_g$ is (in the limit $u_1 \rightarrow 0$)
\begin{equation}
U _g = \left( \begin{array}{cccc} {w_R \over N_g}&0&{u_2^2/w_R
\over N_1}&{0}\\ 0&{w_L \over N_L}&0&{-u_2^2/ w_L \over N_2}\\
{-u_2 \over N_g}&0&{u_2 \over N_1}&0\\ 0&{ u_2 \over N_L}&0&u_2
\over N_2 \end{array} \right) \equiv \pmatrix{U_{11} \ U_{12} \
U_{13} \ U_{14} \cr
         U_{21} \ U_{22} \ U_{23} \ U_{24} \cr
         U_{31} \ U_{32} \ U_{33} \ U_{34} \cr
         U_{41} \ U_{42} \ U_{43} \ U_{44} \cr},
\label{lw2}
\end{equation}
with normalisation constants
\begin{eqnarray}
N_g^2 &=& w_R^2 + u_2^2, \ N_L^2 = w_L^2+u_2^2, \nonumber \\
N_1^2&=& u_2^2 \biggl[ 1 + \left({u_2\over w_R}\right) ^2 \biggr],
\ N_2^2= u_2^2 \biggl[ 1 + \left({u_2\over w_L}\right) ^2 \biggr].
\end{eqnarray}
Referring to Eqs. (\ref{lw1}), (\ref{lw2}), we observe that
\begin{eqnarray}
\chi_R^{4,-{1 \over 2}} &=& {w_R \over \sqrt{w_R^2+u_2^2}}G_R^- +
{u_2 \over \sqrt{w_R^2+u_2^2}}H_{1}^-, \nonumber \\ \chi_L^{4,-{1
\over 2}} &=& {w_L \over \sqrt{w_L^2+u_2^2}}G_L^- - {u_2 \over
\sqrt{w_L^2+u_2^2}}H_{2}^-. \label{u1z}
\end{eqnarray}
Thus it is clear that the $\chi$-loop correction vanishes in the
limit $u_1 \rightarrow 0$. Self consistency implies that
\begin{eqnarray}
\lim_{ {{u_1 \rightarrow 0} \atop {u_2 \ne 0}}} V_{3\langle \phi
\rangle} = \lim_{ {{u_2 \rightarrow 0} \atop {u_1 \ne 0}}}
V_{3\langle \phi \rangle} = 0,
\end{eqnarray}
which we also prove in the Appendix. In short, the $\chi$-loop
contribution to the neutrino Dirac mass vanishes in the limit $u_1
\rightarrow 0$ (or analogously $u_2 \rightarrow 0$). Recall that
the gauge contribution to Dirac mass also vanishes in this limit.
The expected hierarchy $u_1 \ll u_2$ (or $u_2 \ll u_1$) thus
ensures a small gauge {\it and} scalar contribution to the
neutrino Dirac mass.

We now investigate the case of small $u_1 \ne 0$ (with $u_1 \ll
u_2$). In this case we can treat the $u_1$ term as a perturbation
that induces small mass mixing term $V_{3\langle \phi \rangle}$
that will couple $\chi_R^{4,-{1 \over 2}}$ to $\chi_L^{4,-{1 \over
2}}$, thus leading to neutrino Dirac masses. This argument also
holds true if we interchange $u_1 \leftrightarrow u_2$. Indexing
the would-be Goldstone bosons and physical Higgs fields as
\begin{eqnarray}
S_a^T = \pmatrix{G^-_R, G^-_L, H^-_1, H^-_2}, \ a = 1,2,3,4,
\end{eqnarray}
we can write the linear combination of $G^{-}_{L,R}$ and
$H^-_{1,2}$ in $\chi_{L,R}^{4,-{1 \over 2}}$ compactly as
\begin{eqnarray}
\chi_R^{4,-{1 \over 2}} = \sum_{{a' = 1,3}} U_{1a'}S_{a'}, \  \ \
\chi_L^{4,-{1 \over 2}} = \sum_{b=2,4}U_{2b}S_b.
\end{eqnarray}
The explicit expression of $U_{1a'},U_{2b}$ can be read off
directly from $U _g$ in Eq. (\ref{lw2}). The mass correction in
Fig. 3 is now a summation of all the contributions from the
approximate mass eigenstates diagrams. Now we could evaluate the
Feynman diagram in Fig. 3 by treating the cross insertion as a
perturbation that gives rise to a vertex factor $m^2_{3\langle
\phi \rangle} = -(Mu_1 + M'u_2)$. In the special case of decoupled
generations, the mass correction is calculated to be (for first
generation)
\begin{equation}
m_D = k_{\chi }m_{e}, \label{mchi}
\end{equation}
with
\begin{eqnarray}
k_{\chi } &=& \frac{M_{E}^2}{16 \pi^2 w_L w_R}A, \label{kchi1}
\end{eqnarray}
where
\begin{eqnarray}
A &=& \sum_{a',b}A_{a'b}U_{1a'}U_{2b}, \nonumber \\ A_{a'b} &=&
\left({Mu_1 + M'u_2 \over M_{S_{a'}}^2 - M_{E}^2} \right)
\bigg[{M_{S_{a'}}^2 \ln({M_{S_b}^2 \over M_{S_{a'}}^2}) \over
M_{S_b}^2 - M_{S_{a'}}^2} - { M_{E}^2 \ln({M_{S_b}^2 \over
M_{E}^2}) \over M_{S_b}^2 - M_{E}^2}\bigg].
\end{eqnarray}
This quantity $A$ can be greatly simplified in the limit $w_R^2
\gg u_2^2 \gg w_L^2,u_1^2$. In this limit $G^-_R$ and $H^-_2$ will
dominate the loop:
\begin{eqnarray}
A \approx \left({Mu_1 + M'u_2 \over M_{W_R}^2 - M_{E}^2} \right)
\bigg[{M_{H_2}^2 \ln({M_{H_2}^2 \over M_{W_R}^2}) \over M_{H_2}^2
- M_{W_R}^2} - { M_{E}^2 \ln({M_{W_R}^2 \over M_{E}^2}) \over
M_{W_R}^2 - M_{E}^2}\bigg],
\end{eqnarray}
where we have set the mass of $G^\pm_R$ to the mass of $M_{W_R}$
since we are working in the t'Hooft-Feynman gauge. Putting in
reasonable limits for the parameters ($M_{W_R} \stackrel {>}{\sim}
\hbox{0.5 TeV}, 50 \hbox{ GeV} < M_{E} \stackrel {<}{\sim}
\hbox{10 TeV}, 1\hbox{ GeV} <w_{L} < \hbox{200 GeV}$), we find
\begin{eqnarray}
{u_1 u_2 M_E^4 \over w_L w_R M_{W_R}^4} \stackrel {<}{\sim}
k_{\chi} \stackrel {<}{\sim} {u_1 u_2 \over w_L w_R}.
\end{eqnarray}
Thus we can now constrain the relative contribution of the gauge
and $\chi$-loop diagrams. We find that
\begin{eqnarray}
{M_E^4 \over M_{W_R}^4}{w_R \over w_L} \stackrel {<}{\sim}
{k_{\chi} \over k_g} \stackrel {<}{\sim} {w_R \over w_L}.
\end{eqnarray}
Using ${M_{E} \over M_{W_R}} \stackrel {>}{\sim} 10^{-2} $ and $10
\stackrel {<}{\sim} {w_{R} \over w_L} \stackrel {<}{\sim} 10^{4}$,
we have
\begin{eqnarray}
10^{-7} \stackrel {<}{\sim} {k_{\chi} \over k_g} \stackrel
{<}{\sim} 10^4.
\end{eqnarray}
Thus there is a range of parameters where gauge loop dominates
($k_g \gg k_\chi$) and range of parameters where $\chi$-loop
dominates ($k_g \ll k_\chi$). There is also a troublesome
intermediate region with $k_g \approx k_\chi$ which we will for
the most part ignore.

\section{$\tilde \nu_L- \tilde \nu_R$ mixing in decoupled generations}

Having derived the various contributions to the neutral lepton
mass matrix {\bf M}, we now examine the scenario where we assume
that the mixing between generations is approximately negligible.
Referring to the tree level mass matrix $\bf M$ of Eq.
(\ref{yyt2}), $\tilde \nu_R$ gains a Majorana mass from its mixing
with the $E$ leptons while $\tilde \nu_L$ is massless at tree
level. Including the 1-loop mass corrections, the mass matrix $\bf
M$ is
\begin{equation}
\bf M = \left(
\begin{array}{cccc}
m_M&m_D&m_{\nu E}&0\\ m_D^\dagger&0&Y_R M_u Y_L^\dagger U & M_{\em
l}V\\ m_{\nu E}^\dagger& (Y_R M_u Y_L^\dagger
U)^\dagger&0&-M_{E}\\ 0&(M_{\em l}V)^\dagger&-M_{E}&0
\end{array}\right).
\label{yyt}
\end{equation}
The neutrinos (both left-handed and right-handed ones) will
approximately decouple from the $E$ leptons since the latter are
much heavier (recall that $M_{E} \stackrel{>}{\sim} 45$ GeV from
$Z^0$ width). The effective Lagrangian density for the mass matrix
of the neutrinos after decoupling from the $E$ leptons is
\begin{eqnarray}
{\cal L}_{eff} &=& {1 \over 2} \left( \begin{array}{cccc}
\overline {{\tilde \nu}_L} & \overline {({\tilde \nu}_R)^c}
\end{array} \right) M_\nu \left( \begin{array}{cccc} ({\tilde
\nu}_L)^c \\ {\tilde \nu}_R \end{array} \right) + \hbox{ H. c.},
\label{Leff} \end{eqnarray} where the matrix $M_\nu$ is as given
by
\begin{eqnarray}
M_{\nu} \simeq \left(\begin{array}{cccc} m_M& m_D'
\\(m_D')^\dagger&M_R
\end{array}\right),
\label{Mnu}
\end{eqnarray}
with
\begin{equation}
m_D' = m_D + m_{\nu E} M_E^{-1} V^\dagger M_{\em l}.
\label{mdprime}
\end{equation}
$M_R$ is given earlier in Eq. (\ref{gMR}). In the see-saw limit
where the eigenvalues of $M_R$ are much larger than the
eigenvalues of $m'_D$, Eq. (\ref{Leff}) becomes
\begin{equation}
{\cal L}^{see-saw} \simeq {1 \over 2} \overline { {\tilde \nu}_L}
m_L ({\tilde \nu}_L)^c + {1 \over 2} \overline {({\tilde
\nu}_R)^c} M_{R} {\tilde \nu}_R + \hbox{ H.c.}, \label{Lseesaw}
\end{equation}
where
\begin{eqnarray}
m_L  &\simeq& m_M - m_D' M_{R}^{-1}(m_D')^{\dagger} \nonumber \\
 &=& m_M - V_L \left(
\begin{array}{cccc} m_{11}&0&0\\0&m_{22}&0\\0&0&m_{33} \end{array}
\right) V_{R}^\dagger U_{R}^* \left( \begin{array}{cccc}
M_{1}&0&0\\0&M_{2}&0\\0&0&M_{3}\end{array} \right)^{-1} \times
\nonumber \\ && \ \ \ \ \ U_{R}^\dagger V_{R}^* \left(
\begin{array}{cccc} m_{11}&0&0\\0&m_{22}&0\\0&0&m_{33} \end{array}
\right)V_{L}^T, \label{ml}
\end{eqnarray}
with $m_{ii}$ and $M_{i}$ ($i = 1,2,3$) as the eigen masses of the
mass matrices $m_D$ and $M_R$ respectively. In Eq. (\ref{ml}) we
have assumed that the matrices $M_{R}$ and $m_D'$ are diagonalised
with
\begin{equation}
M_{R} = U_{R} \left( \begin{array}{cccc} M_1&0&0\\ 0&M_2
&0\\0&0&M_3 \end{array} \right) U_{R}^{\dagger}, \;\;\; m_D' =
V_{L} \left( \begin{array}{cccc}
m_{11}&0&0\\0&m_{22}&0\\0&0&m_{33} \end{array} \right)
V_{R}^{\dagger}. \label{VLR}
\end{equation}
Generically $m_M$ is always tiny in comparison to the other
contributions to $m_L$ in all generations and shall be dropped
hereafter.

In the case of decoupled generations the mixing angle between
$\tilde \nu_L$ with $\tilde \nu_R$ can be determined from the
matrix $M_\nu$:
\begin{equation}
\tan 2\theta_\nu =  - {2m_D' \over M_R}. \label{tan2nu}
\end{equation}
The eigen masses of $M_\nu$ are
\begin{equation}
{1 \over 2}\left(M_R \pm \sqrt{M_R^2 + 4 m_D'^2} \right) ={M_R
\over 2}\left(1 \pm {1 \over \cos 2\theta_\nu}\right) \label{ml1}
\end{equation}
and the mass squared difference is simply
\begin{equation}
\delta m^2_{\nu} = {M^2_R \over \cos 2\theta_\nu} = \left({2m_{\em
l}m_{q_u} \over M_E}\right)^2{1 \over \cos 2\theta_\nu}.
\label{deltam}
\end{equation}
Under the assumption of decoupled generations, the only way to
solve atmospheric neutrino anomaly is via $\tilde \nu_{\mu L}\to
(\tilde \nu_{\mu R})^c$ oscillations. However, since
experimentally we know that $\delta m^2_{atm} \stackrel{<}{\sim}
10^{-2} \hbox{ eV}^2$, $\sin ^2 2\theta_{atm} \stackrel{>}{\sim}
0.8$, this implies
\begin{equation}
M_E = {2m_{q_u}m_{\em l} \over \sqrt{ \delta m^2_{atm} \cos
2\theta_{atm}}}\stackrel{>}{\sim} 70 \hbox{ TeV} \label {deltame}
\end{equation}
even with the lowest possible fermion masses $m_{q_u} = m_u,
m_{\em l}=m_e$. Thus it is not possible to solve atmospheric
neutrino anomaly with decoupled generations while keeping the
symmetry breaking scale in the interesting range
$\stackrel{<}{\sim}$ few TeV. The above result applies to both
cases where $m_D'$ is dominated by the gauge loop or by the
$\chi$-loop.

\section{Mirror symmetrisation of the alternative 422 model}

As concluded in the last section, decoupled generations cannot
accommodate the atmospheric neutrino anomaly within our assumption
of a low 422 symmetry breaking scale $\stackrel{<}{\sim}$ few TeV.
Nevertheless, we can still have near maximal active-sterile
oscillations at the TeV scale if we mirror symmetrise the
alternative 422 model in the spirit of the Exact Parity Model
(EPM) \cite{EPM}. These models allows parity to be an unbroken
symmetry of nature which it turns out offers a framework to
understand the dark matter, neutrino anomalies and various other
puzzles. (For a review of the evidence see \cite{7 1/2}.) It is
well known that in mirror matter models maximal mixing occurs
naturally because of the unbroken parity symmetry. In the `mirror
symmetric alternative 422 model', $\tilde \nu_L$ will oscillate
maximally to its parity partner $(\tilde \nu_R')^c$. Mirror
symmetrisation can thus provide a mechanism to obtain maximal
mixing whereas the naturally tiny masses (hence mass squared
difference) of the neutrinos are radiatively generated and
naturally small in this TeV scale model.

In the mirror symmetric alternative 422 model, the symmetry group
is extended to $ SU(4)\otimes SU(2)_L \otimes SU(2)_R \otimes
SU(4)'\otimes SU(2)'_L\otimes SU(2)'_R$. Each family has a mirror
partner, which we denote with a prime. Considering $\tilde \nu_L$
and $\tilde \nu_R$, $\tilde \nu_R'$ and $\tilde \nu_L'$ will be
included in the particle content enabling parity to be an unbroken
symmetry. Under the parity transformation, $\tilde \nu_L
\leftrightarrow \gamma_0\tilde \nu_R', \tilde \nu_R
\leftrightarrow \gamma_0\tilde \nu_L'$ (as well as $x \to -x$, of
course). The see-saw Langrangian density of Eq. (\ref{Lseesaw})
becomes
\begin{equation}
{\cal L'}^{see-saw} = {1 \over 2}\pmatrix { \overline {{\tilde
\nu}_L} & \overline {({{\tilde \nu'}_R})^c} & \overline {({{\tilde
\nu}_R})^c} & \overline {{\tilde \nu'}_L} }M'_{\nu} \pmatrix{
({\tilde \nu}_L)^c \cr {\tilde \nu'}_R \cr {\tilde \nu}_R \cr
({\tilde \nu'}_L)^c } + \hbox{ H.c.}, \label{Mnumirror}
\end{equation}
where
\begin{equation}
M'_{\nu} = \pmatrix{ 0&m'&m'_D&0 \cr m'&0&0&m_D' \cr
(m_D')^\dagger & 0 & M_R & 0 \cr 0 & (m_D')^\dagger & 0 & M_R }.
\label{Mnuprime}
\end{equation}
$m'$ is a parity invariant mass mixing term that mixes the
ordinary neutrinos with the mirror neutrinos. In the minimal
mirror matter 422 model, i.e. with only the Higgs fields $\phi,
\chi_L, \chi_R$ and $\phi', \chi_L', \chi_R'$, there is no
mirror-ordinary mixing Yukawa coupling to generate $m'$. This is
quite unlike the case of the EPM model where ordinary-mirror
Yukawa coupling $\overline {(\tilde \nu_L')^c} \phi \tilde \nu_L$
exists because $\tilde \nu_R$ and its parity partner $\tilde
\nu_L'$ are gauge singlets. One could consider the mirror-ordinary
mass mixing term as a dimension-five unrenormalisable bare mass
term $\lambda_0 \overline {f'_R} \phi' f_L \phi / M_h$ so that the
mirror symmetric alternative 422 model becomes effectively a
remnant resulting from the breakdown of some unknown physics at a
much higher scale $M_h$.

Alternatively, the required mirror-ordinary mass mixing term can
be generated if an additional Higgs scalar $\rho^{\alpha \beta,
\alpha' \beta'} \sim (2,2)(2',2')$ exists. The gauge invariant
Yukawa coupling
\begin{equation}
{\cal L}_{\rho} = \lambda_{5} \overline{f_L} \rho \tau_2 {f_R'} +
\lambda_{6} \overline{f_L} \rho^c \tau_2 {f_R'} + \hbox{ H.c.}
\end{equation}
will generate a mirror-ordinary mass mixing term as $\rho$
develops a VEV. Because $\rho$ can couple to $\phi \phi'$ in Higgs
potential it can easily gain a VEV in the right orientation.

To see this more transparently, let's take a look at the part of
Higgs potential containing only the scalar field $\rho$,
\begin{equation}
V(\rho) = M^2_{\rho}\rho^\dagger \rho + m_{\rho} \phi'^\dagger
\phi^\dagger \rho + m'_{\rho}({\rho^c})^\dagger \phi \phi' + {\cal
O}(\rho^3, \rho^4) + \hbox{ H.c}..
\end{equation}
As $\phi, \phi'$ develop VEVs at $\langle\phi\rangle = \langle
\phi'\rangle =\pmatrix{0 & u_2 \cr u_1 & 0}$, the trilinear term
will induce a linear term in $\rho$. Because of this, minimising
$V(\rho)$ with respect to $\rho^\dagger$
\begin{eqnarray}
{\partial V \over \partial \rho^\dagger} = M_\rho^2 \rho + m_\rho
\langle\phi \rangle \langle\phi' \rangle + m_\rho' \langle \phi'^c
\rangle \langle \phi^c\rangle = 0
\end{eqnarray}
will induce the VEVs
\begin{equation}
\langle \rho\rangle = -{m_{\rho}\langle \phi\rangle \langle
\phi'\rangle +m'_{\rho}\langle \phi'^c\rangle \langle\phi^c
\rangle \over M_{\rho}^2}
\end{equation}
to the components of $\rho^{\alpha \beta, \alpha' \beta'}$.  The
VEV for the component \begin{eqnarray}
\langle\rho\rangle^{(12,2'1')} = \langle \rho^c
\rangle^{(12,2'1')}= -{(m_{\rho}+m'_{\rho})u_1u_2 \over
M_{\rho}^2}
\end{eqnarray}
will generate the desired ordinary-mirror mass mixing term in
${\cal L}_{\rho}$:
\begin{equation}
{\cal L}_{\rho} = \overline{\tilde\nu_L}m'\tilde\nu_R' + \hbox{
H.c.},
\end{equation}
where
\begin{equation}
m' = -\lambda_5 \langle\rho \rangle^{(12,2'1')} - \lambda_6
\langle\rho^c \rangle^{(12,2'1')} = {u_1 u_2 (m_\rho+m_\rho')\over
M_\rho^2}(\lambda_5 + \lambda_6).
\end{equation}
In any case the mirror-ordinary mass mixing term $m'$ is still a
free parameter of the theory. In the see-saw limit ${\tilde
\nu}_L, ({\tilde \nu_R}')^c$ decouple from ${\tilde \nu}_L',
({\tilde \nu_R})^c$ in Eq. (\ref{Mnumirror}) and the Lagrangian
density of light neutrinos is given by
\begin{equation}
{\cal L}^{''}_{\nu} = {1 \over 2}\pmatrix { \overline{{\tilde
\nu}_L} & \overline {({\tilde \nu}'_R)^c}} M^{''} \pmatrix{
({\tilde \nu}_L)^c \cr {\tilde \nu'}_R } + \hbox{ H.c.},
\label{Mnumirror2}
\end{equation}
where
\begin{equation}
M^{''} = \pmatrix{-m_{D}'M_R^{-1}(m_{D}')^\dagger & m' \cr m' &
-m_{D}'M_R^{-1}(m_{D}')^\dagger}. \label{maxM}
\end{equation}
The mass matrix $M^{''}$ in Eq. (\ref{maxM}) describes maximal
${\tilde \nu}_L \to ({\tilde \nu'}_R)^c$ oscillations with eigen
masses
\begin{equation}
m_\pm = -m_D'M_R^{-1}(m_D')^\dagger \pm m'.
\end{equation}
In the limit of small intergenerational mixing the mass squared
difference is simply
\begin{equation}
\delta m_{\mp}^2 = {4m'm_D'^2 \over M_R} = {2m'm_{\em l}M_{E_i}
S^2\eta^2 \over m_{q_u}}.
\end{equation}
$\delta m_{\pm}^2$ can be identified as $\delta m^2 _{atm}$
(second generation) and $\delta m^2_{solar}$ (first generation)
with maximal oscillation solutions, thereby explaining the atmospheric
and solar neutrino anomalies. If small intergenerational mixing
(parametrised by $\theta$ and $\phi$) between the first and second
generations is included, the LSND experiment $\delta m_{L \! S \!
N \! D}^2$ can be identified with
\begin{eqnarray}
\delta m_{L \! S \! N \! D}^2 \equiv |\delta m_{\nu_{e+},\nu_{\mu
+}}^2| &\equiv& |m_{\nu_{e+}}^2 - m_{\nu_{\mu +}}^2|
\end{eqnarray}
and
\begin{eqnarray}
(\sin 2\theta + \sin 2\phi)^2 \sim 3 \times 10^{-2} - 10^{-3}.
\end{eqnarray}
Thus, the mirror symmetrised extension of the alternative 422
model can explain all the three neutrino anomalies without any
physics beyond the TeV scale. The alternative 422 model provides
the neutrino masses while mirror symmetrising it provides the
maximal mixing between each ordinary and mirror neutrino flavour.
\section{Mixing between generations: neutrino mass dominated by
gauge interactions}

In the previous two sections we have examined the case where the
mixing between generations was small. Under this assumption the
minimal alternative 422 model could not accommodate the large
mixing required to explain the atmospheric or solar neutrino
problems. However we also shown that the mirror symmetrised
extension could explain all three neutrino anomalies. We now
return to the minimal alternative 422 model and examine the
alternative case where mixing between generations is large. From
the discussion in the previous sections, recall that there are two
different ways in which neutrino masses are generated in the
model, namely via the gauge interactions and via the (scalar)
$\chi$-interactions. In either case, we see that the Dirac masses
are proportional to the charged lepton masses $m_{\em l}$ [see Eq.
(\ref{md}), (\ref{mchi})]. The proportional constants $k_{\chi}$
and $k_g$ are poorly constrained however. Our ignorance of their
relative strength does not permit us to tell which mechanism is
the dominating one. Though it remains a possibility that both
contributions may be equally contributive, we shall not treat this
general scenario to avoid complications. We will focus our
attention only to the limiting case where the gauge interactions
are assumed to dominate over the scalar interactions, i.e. $k_g
\gg k_\chi$.

\subsection{Two generations maximal mixing and `lop-sided' $M_R$}

In this subsection, we shall analyse  the limiting case where the
radiative contribution to the neutrino mass from gauge
interactions dominates the contribution due to the scalar
interactions, i.e. $k_g \gg k_\chi$. Our strategy is to look for a
mechanism in the 2-3 sector that provides a near maximal
$\tilde\nu_{\mu L} \to \tilde\nu_{\tau L}$ oscillation solution to
atmospheric neutrino anomaly. We then generalise it to the case of
three generations.

Recall that the effective mass matrix for the light neutrinos is
given by $m_L$ [see Eq. (\ref{ml})]. Knowledge of $m_L$ allows us
to work out the mixing angles and $\delta m^2$, and to make
contact with the neutrino experiments. In general, to calculate
$m_L$ we require knowledge of the Yukawa coupling matrices
$\lambda_1, ... ,\lambda_4$. Our purpose here is to identify
simple forms of $\lambda_1,...,\lambda_4$ which enable the model
to accommodate the neutrino data. Let us first look at the special
case of just the 2-3 sector. As discussed in Ref. \cite{paper1},
we can obtain maximal $\tilde\nu_{\mu L} \to \tilde\nu_{\tau L}$
oscillations if $m_D'$ is approximately diagonal and $M_{R}$ being
`lop sided', meaning that the off diagonal elements are much
larger than the diagonal elements of the 2-3 mass matrix.

The (tree level) two generations mass matrix $M_{R}$ in the 2-3
sector, which we parametrise by
\begin{equation} M_{R2} = \left( \begin{array}{cccc} r_{22}&
r_{23}\\ r_{23}& r_{33}\end{array}\right)
\end{equation}
can be easily worked out from Eq. (\ref{gMR}). Note that the form
of $M_{R2}$ is intimately related to the Yukawa couplings
$\lambda_1$ as well as the right-handed CKM-type matrix $Y_R$. On
the other hand, quite independently, the radiative correction to
the neutrino mass $m_D'$ can be read off directly from Eq.
(\ref{mdprime}). Cross generational mixing of the Dirac masses is
possible due to the dependence of the matrix $V$. However, in the
limiting case of diagonal $V$, the matrix $m_{\nu E}$, and thus
$m_D'$, also becomes diagonal. In this case we can approximate
$m_D'$ by $m_D$ \footnote{In the case where the radiative
correction is dominated by the gauge interactions $m_{\nu E}$ will
contribute to $m_D'$ [see Eq. (\ref{mdprime})]. In the special
case of decoupled generations, $m_D' = m_D + m_{\nu E} m_{\em l} /
M_{E_i}$. Ref. \cite{422.2} established the estimation of $m_D
\sim {m_{\nu E} m_{\em l} / M_{E_i}}.$ Assuming no accidental
cancellation, we could approximate $m_D' \sim m_D$.}. The Dirac
masses of $m_D' \sim m_D$, namely $m_{22}, m_{33}$, are simply
given by Eq. (\ref{mddcpl}), with $m_{\em l} = m_{\mu}, m_{\tau}$
respectively.

Let us parametrise the unitary matrices [as introduced in Eq.
(\ref{basis})] as
\begin{eqnarray} U = \pmatrix{\cos \theta_u &-\sin\theta_u \cr
\sin \theta_u & \cos \theta_u}, & V = \pmatrix{\cos \theta_v &
-\sin \theta_v \cr \sin \theta_v & \cos \theta_v}, \ Y_R =
\pmatrix{\cos \theta_y & -\sin \theta_y \cr \sin \theta_y & \cos
\theta_y}.
\end{eqnarray}
The simplest way that we have found to obtain lop-sided $M_R$ in
our scheme (modulo certain `permutations' which we will discuss
later) is to note that when $\lambda_1$ is approximately diagonal
(i.e. $\theta_{u,v}$ small)\footnote{ For simplicity we have set
the CKM matrix $Y_L = \bf I$.}, then
\begin{eqnarray}
\lim_{\theta_{u,v} \rightarrow 0} M_{R2} &=& \left(
\begin{array}{cccc} {2\cos\theta_y m_\mu m_c \over M_{E_2}} &
{\sin\theta_ym_\mu m_c \over M_{E_2}} - {m_\tau m_t \over M_{E_3}}
\\ {\sin \theta_y m_\mu m_c \over M_{E_2}} - {m_\tau m_t \over
M_{E_3}} & {2\cos\theta_y m_\tau m_t \over M_{E_3}}
\end{array}\right). \end{eqnarray}
If we let $M_{R2}$ be diagonalised by the unitary matrix
\begin{equation} U_{R2} = \left(
\begin{array}{cccc}\cos \theta_R&\ -\sin \theta_R\\\sin
\theta_R&\cos \theta_R \end{array} \right),\label{thetaR}
\end{equation}
we find that the mixing angle $\theta_R$ behaves like
\begin{eqnarray} \lim_{\theta_{u,v} \rightarrow 0} \tan 2\theta_R
&=&-\frac{m_\tau m_t / M_{E_3}}{( {m_\tau m_\mu \over M_{E_2}}   -
{m_\tau m_t \over M_{E_3}})\cos\theta_y}. \label{limitmr}
\end{eqnarray} This mean that, assuming no accidental cancellation,
we will obtain $|\tan 2 \theta_R| \gg 1$ in the limit
\begin{equation} \theta_{u,v} \rightarrow 0,
\ \theta_y \rightarrow {\pi \over 2}, \label{ansatzy}
\end{equation} with a pair of approximately  degenerate eigen masses
\begin{equation}
M_{2,3} \simeq \mp \left( {m_\mu m_c \over M_{E_2}} - {m_\tau m_t
\over M_{E_3}} \right) \equiv \mp m_{\nu_R}. \label{M23}
\end{equation}
Since $m_D'$ is kept diagonal due to the vanishing $\theta_v$,
therefore maximal mixing in the left-handed neutrinos (2-3 sector)
is realised. This can be seen from the effective mass matrix $m_L$
of Eq. (\ref{ml}):
\begin{equation}
m_L \rightarrow  \left( \begin{array}{cccc} 0&1\\1&0\end{array}
\right) \frac{m_{22}m_{33}}{m_{\nu_R}}. \label{yqw}
\end{equation}
The effective mass matrix Eq. (\ref{yqw}) indicates that, in the
presence of small intergenerational mixing, the left-handed active
neutrinos $\tilde\nu_{\mu L}, \tilde\nu_{\tau L}$ are approximate
maximal mixture of almost degenerate mass eigenstates. We denote
these eigen masses as
\begin{equation}
m_2, m_3 = {\pm \eta^2 S^2 \over ({m_c \over m_\tau M_{E_2}} -
{m_t \over m_\mu M_{E_3}})}. \label{mll}
\end{equation}

In the case where the limits of Eq. (\ref{ansatzy}) are only
approximate, the diagonal entries in $M_{R2}$ (and hence $m_L$)
shall in effect be replaced by some tiny values $\epsilon$
instead. It is natural to conceive that $\epsilon$ could split the
eigen mass degeneracy to an order of $10^{-3}\hbox{ eV}^2
\stackrel{<}{\sim} |\delta m_{atm}^2| \equiv |m_{3}^2 - m_{2}^2|
\stackrel{<}{\sim} 10^{-2} \hbox{ eV}^2$. However, if this
splitting is to be natural, the size of $\delta m^2_{3,2} \equiv
m^2_{3} - m^2_{2}$ must be smaller than $m^2_2, m^2_3$ themselves,
i.e.,
\begin{eqnarray}
m^2_{3},m^2_{2} \gg 10^{-2} \hbox{ eV}^2.\label{m23req}
\end{eqnarray}
Estimating $m^2_{2,3}$ [using the range of $S$ from Eq.
(\ref{Srange})],
\begin{eqnarray}
&&m^2_{2,3} \approx M^2_{E_3}\bigg( {m_\mu \over m_t}\bigg)^2 S^4
\eta ^4 \stackrel{<}{\sim} 10^{-7} \bigg( {M_{E_3} \over
\hbox{TeV}} \bigg)^2 \hbox{ eV}^2.
\end{eqnarray}
Thus, we see that although near maximal
$\tilde\nu_{\mu L} \to \tilde\nu_{\tau L}$ oscillations are achieved via the
ansatz of Eq. (\ref{ansatzy}) the $\delta m^2$ is not in natural
compatibility with $\delta m^2_{atm}$ in this particular
situation. This means that maximal oscillation solution to
atmospheric neutrino anomaly is not accommodated in this case.
However we will show later that a permutation on the up-type quark
masses in $m_i$ could be performed to obtain compatibility with
the consistency requirement of Eq. (\ref{m23req}). It is more
convenient to discuss how the permutation (and its rationale) of
the up-type quark masses could obtain a compatible range of
$\delta m^2$ between this scheme and the experimental values by
including the first generation neutrino into the picture. We shall
do so in the following subsection.

Before we proceed to the three generations case, it is worth to
comment on the form taken by $Y_R^\dagger$ under the ansatz of Eq.
(\ref{ansatzy}). Since $m_D$ is approximately diagonal under this
ansatz, near maximal $\tilde\nu_{\mu L} \to \tilde\nu_{\tau L}$
oscillations should originate from $M_{R2}$ of the right-handed
neutrino sector, which is in turn related to the form of
$Y_R^\dagger$ and the mass matrix $\lambda_1$. In the limit of
this ansatz, $Y_R^\dagger$ is off-diagonal in the 2-3 sector
(assuming that the first generation is approximately decoupled
from the 2-3 sector),
\begin{equation} Y_R^\dagger \sim \pmatrix{1&0&0 \cr 0&0&1 \cr
0&1&0}. \label{yrf}
\end{equation}
If we assume a left-right similarity so that
\begin{eqnarray}
K_R \sim K_L \sim {\bf I}, \label{commonorg}
\end{eqnarray}
then this means that $K' \equiv Y_R^\dagger K_R^\dagger$ also has
the form
\begin{equation}
K' \sim \pmatrix{1&0&0 \cr 0&0&1 \cr 0&1&0} = Y^\dagger.
\label{Kf1}
\end{equation}
Recall that, as discussed in detail by Ref. \cite{422.1,422.2},
the $SU(4)$ gauge interactions involving the coloured gauge bosons
$W_\mu'$
\begin{equation}
{\cal L} = {g_s \over \sqrt 2}\overline{D_R}W_\mu' \gamma^\mu
K'{\em l}_R + \hbox{ H. c.}
\end{equation}
could mediate lepto-quark transitions. In Ref. \cite{422.1,422.2}
it was shown that $K'$ must be non-diagonal to avoid contributions
from $K^0 \to \mu^\pm e^\mp$ decays. In that case the primary
constraint on the $SU(4)$ symmetry breaking scale $M_{W'}$ is from
rare $B_0$ decays, $B_0 \rightarrow \mu^\pm e^\mp, \tau^\pm e^\mp
\hbox{ and } \tau^\pm \mu^\mp$, depending on the forms of $K'$,
resulting in the low symmetry breaking scale of $\sim$ TeV.
Specifically, in order for the TeV symmetry breaking scale to
occur, $K'$ have to be in certain forms that will suppress the
rare decays $K_L \rightarrow \mu^\pm e^\mp$. In fact Ref. \cite
{422.2} pointed that there are only 4 possible (approximate) forms
for $K'$ that are consistent with the TeV SSB scale:
\begin{eqnarray}
K'_1 = \pmatrix{0 & 0 & 1 \cr \cos \alpha & \sin \alpha & 0 \cr
-\sin \alpha & \cos \alpha & 0}, \ \ K'_2 = \pmatrix{\cos \beta &
\sin \beta & 0 \cr 0 & 0 & 1 \cr -\sin \beta & \cos \beta & 0},
\nonumber \\ K'_3 = \pmatrix{\cos \gamma & 0 &\sin \gamma \cr
-\sin \gamma & 0 & \cos \gamma \cr 0 & 1 & 0}, \ \ K'_4 =
\pmatrix{ 0 & \cos \delta & \sin \delta \cr 0 & -\sin \delta &
\cos \delta \cr 1&0&0}. \label{Kprime}
\end{eqnarray}
Note that Eq. (\ref{Kf1}) is a special case of these forms,
namely,
\begin{eqnarray}
K'_2(\beta = 0) = K'_3(\gamma = 0). \label{K23}
\end{eqnarray}
Eq. (\ref{commonorg}) amounts to suggesting that non-diagonal $K'$
(required for low symmetry breaking) and non-diagonal
$Y_R^\dagger$ (required to obtain maximal $\tilde\nu_{\mu L} \to \tilde\nu_{\tau L}$ oscillations) may have a common
origin. For example, if we make the ansatz that $\lambda_3,
\lambda_4$ takes the approximate form
\begin{eqnarray} \lambda_3
\sim \pmatrix{\times &0&0 \cr 0&0&\times \cr 0&\times&0}, \
\lambda_4 \sim \pmatrix{\times&0&0 \cr 0&0&\times \cr 0&\times&0},
\label{texture34}
\end{eqnarray}
with the other 2 Yukawas $\{\lambda_{1}, \lambda_{2}\}$
approximately diagonal, then this will simultaneously lead to Eq.
({\ref{yrf}) and ({\ref{Kf1}}), hence the low symmetry breaking
scale ($\sim$ few TeV) and near maximal
$\tilde\nu_{\mu L} \to \tilde\nu_{\tau L}$ oscillations.

\subsection{Permutation of up-type quark masses and three
generations mixing}

In this subsection we will include the first generation neutrino
into the picture under the assumption that the intergenerational
mixing between the first and the 2-3 sector is small. In such a
scenario $m_L$ will take the approximate form
\begin{equation}
m_L \approx \left( \begin{array}{cccc} -\frac{m_{11}^2}{M_1}
&0&0\\ 0&0&\frac{m_{22}m_{33}}{m_{\nu_{R}}}\\
0&\frac{m_{22}m_{33}}{m_{\nu_{R}}}&0 \end{array}\right)
\label{m111}
\end{equation}
which leads to a pair of almost degenerate eigen masses $m_{2},
m_{3}$ in the 2-3 sector that is separated from eigen mass
\begin{eqnarray}
m_{1}= -\frac{m_{11}^2}{M_1} = - {m_e M_{E_1}\eta^2 S^2\over 2
m_u} \hbox{ eV} \label{mll1}
\end{eqnarray}
with a distinct gap. The near maximal oscillation solution for
atmospheric neutrino anomaly corresponds to a near degenerate pair
of eigen masses $m_{2},m_{3}$. It is obvious that to include the
first generation neutrino to solve the solar neutrino anomaly
(which involves a much smaller $\delta m^2 \stackrel{<}{\sim}
10^{-3} \hbox{ eV}^2$ required by all oscillation solutions) the
first generation neutrino mass will have to be very nearly
degenerate with the other two neutrino masses, which is quite
unnatural. Thus we conclude that the gauge loop mechanism seems to
explain the LSND data more readily than the solar neutrino
anomaly. The gap $\delta m_{1,3}^2 \approx \delta m^2_{1,2}$ can
then be identified with the LSND measurement, $0.2\hbox{ eV}^2
\stackrel{<}{\sim} \delta m_{L\!S\!N\!D}^2 \stackrel{<}{\sim}
3\hbox{ eV}^2$ \footnote{Note that the solar neutrino problem can
be solved in the current scenario if we mirror symmetrize the
model as we did in section VI so that maximal $\nu_e \to \nu_e'$
oscillations result. In particular if we want to explore the
possibility that the atmospheric neutrino anomaly is solved via
near maximal $\nu_\mu \to \nu_\tau$ oscillations then this is
actually {\it compatible} with mirror symmetry, since we just need
to be in the parameter region where the oscillation length for
$\nu_\mu \to \nu'_\mu$ oscillations is much greater than the
diameter of the earth for atmospheric neutrino energies.}.

In the case of two generations discussed in the previous
subsection, the form of the matrix $K' \equiv K'_2(\beta = 0) =
K'_3(\gamma = 0)$ in Eq. (\ref{Kf1}) leads to the scale of
$m^2_{2,3}$ that are incompatible to Eq. (\ref{m23req}). However,
this result corresponds to only one specific form of $K'$ in Eq.
(\ref{Kprime}). In general, the other forms of $K'$ in Eq.
(\ref{Kprime}) could also lead to maximal neutrino oscillations
that are consistent with TeV scale SSB, assuming that $Y_R^\dagger
= K'$ holds. Essentially there are only 4 special forms of $K'$
that are of our interest. The case of $K' \equiv K'_2(\beta = 0) =
K'_3(\gamma = 0)$ has been shown to be incompatible with Eq.
(\ref{m23req}). We will investigate the other three forms of $K'$
in turns, namely $(a) \ K'_4(\delta = 0) = K'_2(\beta = {\pi \over
2})$, $(b) \ K'_1(\alpha = 0) = K'_3(\gamma = {\pi \over 2})$ and
$(c) \ K'_4(\alpha = {\pi \over 2}) = K'_4(\delta = {\pi \over
2})$.

To find out how these three different forms of $K'$ can also lead
to maximal mixing in the 2-3 sector, we will use the constraint
that the lop-sided form of $M_R$, and thus maximal $\tilde\nu_{\mu
L} \to \tilde\nu_{\tau L}$ oscillations, is preserved when
$Y_R^\dagger$ takes on different form other than that of Eq.
(\ref{yrf}) as we replace $Y_R^\dagger \rightarrow Y_R'^\dagger$.
[$Y_R'^\dagger$ are the other forms of $K'$ as catalogued in Eq.
(\ref{Kprime})].

Referring to $M_R$ in Eq. (\ref{gMR}), since we know that, with
the ansatz of Eq. (\ref{ansatzy}) and the choice of basis, the
matrices $M_{ \em l}, V, U^\dagger, M_E, Y_L$ are diagonal, it
then follows that the form of $M_R$ goes like
\begin{eqnarray}M_R \sim U^\dagger M_u Y_R^\dagger \sim
\pmatrix{\times & 0 & 0\cr0 & 0 & \times \cr 0 &\times
&0}.\end{eqnarray} When $Y_R^\dagger$ is replaced by
$Y_R'^\dagger$ via some transformation $T$,
\begin{eqnarray}
Y_R^\dagger\rightarrow Y_R'^\dagger = T Y_R^\dagger, \label {YRT}
\end{eqnarray}
the lop-sided form of $M_R$ should be preserved,
\begin{eqnarray} U^\dagger M_u Y_R^\dagger \rightarrow U'^\dagger
M_u T Y_R^\dagger\sim \pmatrix{\times & 0 & 0\cr0 & 0 & \times \cr
0 &\times &0}.
\end{eqnarray} Since $Y_R^\dagger = \pmatrix{1 & 0 & 0\cr0 & 0 & 1
\cr 0 &1 &0}$ [see Eq. (\ref{yrf})], this mean that $U'^\dagger$
ought to conform to the condition that
\begin{eqnarray}
U'^\dagger M_u T = \pmatrix{\times & 0 & 0\cr0 & \times & 0 \cr 0
&0 &\times}. \label{UT}
\end{eqnarray}
The matrix $T$ relates the `original' form of $Y_R^\dagger$ [as in
Eq. (\ref{yrf})] to $Y_R'^\dagger$ via Eq. (\ref{YRT}). Once we
know $T$ then we could work out $U'^\dagger$ from Eq. (\ref{UT}).
With $Y_R^\dagger \rightarrow Y_R'^\dagger$ and $U^\dagger (= {\bf
I}) \rightarrow U'^\dagger$, the net effect is that the diagonal
up-type quark mass matrix $M_u$ is replaced by
\begin{eqnarray}
M_u \rightarrow U'^\dagger M_u T,
\end{eqnarray}
thus changing the up-type quark masses dependence of the eigen
masses $m_{i}$.

\vskip 0.4cm \noindent $(a)$ \underline{{$ Y_R'^\dagger \equiv
K'_4(\delta = 0) = K'_2(\beta = {\pi \over 2})$}}:

\noindent Let's look at the form of $Y_R'^\dagger \equiv
K'_4(\delta = 0) = K'_2(\beta = {\pi \over 2}) = \pmatrix{0 & 1 &
0 \cr 0 & 0 & 1 \cr 1& 0 & 0}$. The corresponding $T$ matrix is $T
= \pmatrix{0 & 0 & 1 \cr 0 & 1 & 0 \cr 1& 0 & 0}$. The matrix
$U'^\dagger$, by Eq. (\ref{UT}), is $U'^\dagger = \pmatrix{0 & 0 &
1 \cr 0 & 1 & 0 \cr 1& 0 & 0}$,
\begin{eqnarray}
\Rightarrow M_u = \hbox{ diag } \{ m_u, m_c, m_t \} \rightarrow
U'^\dagger M_u T = \hbox{ diag }\{ m_t, m_c, m_u \}.
\label{permute1}
\end{eqnarray}
As a result, the up-type quark masses as appear in $m_{2,3}$ and
$m_1$ in Eq. (\ref{mll}) and (\ref{mll1}) will be permuted by Eq.
(\ref{permute1}) which lead to the eigen masses for the light
neutrinos
\begin{eqnarray}
m_{2},m_{3} \approx \mp \eta^2 S^2 M_{E_2}\left({m_\tau \over
m_c}\right), \ m_{1} = -{\eta^2 S^2 M_{E_1} \over 2}\left({m_e
\over m_t}\right).
\end{eqnarray}
We see that the upper limit of the scale of $m^2_{2,3}$ are of the
order
\begin{equation}\label{m23req1}
m^2_{2,3}/\hbox{eV}^2 \stackrel{<}{\sim} 200\eta^4,
\end{equation}
which permits a large range of parameter in $\eta$ so that
$m^2_{2,3} \gg \delta m^2_{atm}$ for self consistency. The mass
gap
\begin{eqnarray} |m^2_{2} - m^2_{1}| \approx m^2_{2} = {S^4\eta^4}
M^2_{E_2}\left({m_\tau \over m_c}\right)^2 \label{lsnd2}
\end{eqnarray}
also permits a large range of parameter space in $\eta$ to
accommodate $\delta m^2_{L\!S\!N\!D}$. This scheme is easily
realised by imposing the ansatz that the Yukawas are of the forms
\begin{eqnarray} \lambda_2 = \pmatrix {\times &0&
0 \cr 0& \times &0 \cr  0& 0 & \times}, \ \lambda_1 = \pmatrix
{0&0& \times \cr 0& \times &0 \cr  \times& 0 & 0}, \ \lambda_{3,4}
= \pmatrix{0 &\times &0 \cr 0 & 0 & \times \cr \times & 0
&0}.\end{eqnarray}

\vskip 0.4 cm \noindent $(b)$ \underline{{$Y_R'^\dagger \equiv
K'_1(\alpha = 0) =K'_3(\gamma = {\pi \over 2})$}}:

\noindent Next, we look at the form $Y_R'^\dagger \equiv
K'_1(\alpha = 0) =K'_3(\gamma = {\pi \over 2}) = \pmatrix{0&0&1\cr
1&0&0\cr0& 1&0}$. In this case, $U'^{\dagger}$ and $T$ assume the
forms
\begin{equation}
U^\dagger = \pmatrix{0 & 1 & 0 \cr 1 & 0 & 0 \cr 0& 0 & 1}, \ \ T=
\pmatrix{0 & 1 & 0 \cr 1 & 0 & 0 \cr 0& 0 & 1}.
\end{equation}
This results in permutation of up-type quark masses
$\{m_u,m_c,m_t\} \rightarrow \{m_c,m_u,m_t\}$. The mass square
difference \begin{eqnarray}|m^2_{2} - m^2_{1}| \approx m^2_{2} =
{S^4\eta^4} M^2_{E_3}\left({m_\mu \over m_t}\right)^2
\stackrel{<}{\sim} 10^{-7}\bigg( {M_{E_3}\over \hbox{TeV}}\bigg)^2
\hbox { eV}^2 \label{lsnd3}
\end{eqnarray}
is not compatible with
the experimental values of $\delta m^2_{L\!S\!N\!D}$. In addition,
the scale of $m^2_{2,3}$ is also too tiny to accommodate $\delta
m^2_{atm}$:
\begin{equation}\label{m23req2}
m^2_{2,3} < 4 \times 10^{-5} \hbox{ eV}^2.
\end{equation}
In passing, we note that the Yukawas of the forms
\begin{eqnarray}
\lambda_2 = \pmatrix {\times &0& 0 \cr 0& \times &0 \cr  0& 0 &
\times}, \ \lambda_1 = \pmatrix {0&\times& 0 \cr \times& 0 &0 \cr
0& 0 & \times}, \ \lambda_{3,4} = \pmatrix{0 &0&\times \cr \times
& 0 & 0 \cr 0&\times & 0}\end{eqnarray} will lead to the above
unsatisfactory scheme.

\vskip 0.4cm \noindent $(c)$ \underline{$Y_R'^\dagger \equiv
K'_1(\alpha = {\pi \over 2 }) = K_4'(\delta = {\pi \over 2})$}:

\noindent Finally, we investigate the case $Y_R'^\dagger \equiv
K'_1(\alpha = {\pi \over 2 }) = K_4'(\delta = {\pi \over 2}) =
\pmatrix{0&0&1\cr 0&1&0\cr1& 0&0}$. In this case, $U'^{\dagger}$
and $T$ assume the forms
\begin{equation}
U'^\dagger = \pmatrix{0 & 0 & 1 \cr 1 & 0 & 0 \cr 0& 1 & 0}, \ \
T= \pmatrix{0 & 1 & 0 \cr 0 & 0 & 1 \cr 1& 0 & 0}.
\end{equation}
The up-quark mass permutation is $\{m_u,m_c,m_t\} \rightarrow
\{m_t,m_u,m_c\}$. The upper bound of the scale of $m^2_{2,3}$ is
\begin{equation}\label{m23req3}
m^2_{2,3}/\hbox{eV}^2 \stackrel{<}{\sim} 0.7\eta^4,
\end{equation}
which means that it is possible to accommodate $\delta m^2_{atm}$.
The mass squared difference for LSND is
\begin{eqnarray}
\delta m^2_{L\!S\!N\!D} \equiv |m^2_{2} - m^2_{1}| \approx m^2_{2}
= {S^4\eta^4} M^2_{E_3}\left({m_\mu \over m_c}\right)^2
\stackrel{<}{\sim} 10^{-2}\bigg({M_{E_3}\over \hbox{TeV}}\bigg)^2
\hbox{eV}^2. \label{lsnd4}
\end{eqnarray}
In this case we see that its parameter space can still accommodate
the LSND and atmospheric neutrino anomalies [although the regime
of parameter space is more restricted compared to the case $(a)$].
This scheme can be implemented if the Yukawas take the forms
\begin{eqnarray}
\lambda_2 = \pmatrix {\times &0& 0 \cr 0& \times &0 \cr  0& 0 &
\times}, \ \lambda_1 = \pmatrix {0&0& \times \cr 0 & \times&0 \cr
\times& 0 &0}, \ \lambda_{3,4} = \pmatrix{0 &\times&0 \cr 0 & 0 &
\times \cr \times&0 & 0}.\end{eqnarray}

In short, we see that out of the 4 different forms of $K'$ in Eq.
(\ref{Kprime}), the schemes $(a)$ and $(c)$ stand out to be most
promising to provide viable solutions to both LSND and (near
maximal oscillations) atmospheric neutrino anomaly with mass scale
$\stackrel{<}{\sim}$ few TeV. Note the schemes we have discussed
in this section cannot accommodate solar neutrino anomaly (while
keeping both LSND and atmospheric neutrino solutions) because the
right-handed neutrinos are decoupled from the left-handed ones at
TeV scale.


In the parameter range where scalar interactions dominate over the
gauge interactions things are less constrained and there are more
possibilities. It is possible to implement the lop-sided $M_R$
scheme in the scalar sector case in much the same way as in the
previous section. Other possibilities where $m_D$ is off diagonal
is also possible in the scalar case which can also lead to schemes
compatible with data. However these schemes seem less elegant
because of the larger degree of arbitrariness in scalar
interactions.

\section{Conclusion}
The similarity of the quarks and leptons suggests that quarks and
leptons might be connected by some spontaneously broken symmetry.
However such a situation will lead to a gauge hierarchy problem
unless the symmetry breaking scale is less that a few TeV. There
are only two known ways that quark-lepton unification can occur at
the TeV scale. First, quarks and leptons can be connected by a
spontaneously broken discrete symmetry \cite{QLS}. While this an
interesting possibility, it is difficult to naturally explain the
lightness of the neutrinos in these schemes. The second
possibility is a modification of the Pati-Salam model \cite {ps}
called the alternative 422 model \cite{422.1,422.2}. It turns out
that the neutrinos in the alternative 422 model are naturally
light because they are massless at the tree level and their masses
are radiatively generated. The model also predicts novel $B^o$
physics.

The possibility that the model can provide the interactions to
generate the right neutrino mass and mixing patterns which might
explain the atmospheric, solar and/or LSND neutrino anomalies has
been studied in detail in this paper. We have shown that the model
cannot accommodate simultaneously all three of the anomalies
unless it is extended in some way. However the minimal model can
quite naturally accommodate the atmospheric and LSND anomalies. We
have also pointed out that the solar neutrino problem could be
most naturally explained if the model was extended with a mirror
sector.

\vskip 0.8cm \noindent {\bf Acknowledgements} \vskip 0.4cm
\noindent R.Foot is an Australian Research Fellow.\\ T.L.Yoon is
supported by OPRS and MRS from The University of Melbourne.

\newpage

\centerline{\Large \bf Appendix}

\vskip 3 mm

In this Appendix we will obtain the precise form of the would-be
Goldstone bosons $G^\pm_{L,R}$ and physical Higgs fields
$H^\pm_{1,2}$ in terms of the charged weak eigenstate fields
$\chi^{4,-{1 \over 2}}_{L,R}$, $\phi^{{1 \over 2},{1 \over 2}}$
and $\phi^{-{1 \over 2},-{1 \over 2}}$ [i. e. the matrix $U_g$ in
the Eq. (\ref{lw1})]. In addition we will also obtain the
interesting result that
\begin{eqnarray}
\lim_{ {{u_1 \rightarrow 0} \atop {u_2 \ne 0}}} V_{3\langle \phi
\rangle} = \lim_{ {{u_2 \rightarrow 0} \atop {u_1 \ne 0}}}
V_{3\langle \phi \rangle} = 0.
\end{eqnarray}

The Goldstone bosons associated with each spontaneously broken
symmetry are given by
\begin{equation}
G = \Phi^T T^a \lambda_a,
\end{equation}
where $\Phi$ are the Higgs scalar of the theory, $T^a$ the
generators of the broken symmetries and $\lambda_a$ the vacua.
$G^\pm_{L}$ is the would-be Goldstone bosons that are eaten by
$W^{\pm}_{L}$ as their longitudinal polarisation. The associated
(charged) generators of the broken $SU(2)_L$ are $\tau^{\pm}_L$.
Generally,
\begin{eqnarray}G^\pm_{L}& = & \bigg[ \chi_{L}^T
\tau^{\pm}_L \langle \chi_{L} \rangle + \chi_{L}^{cT} \tau^{\pm}_L
\langle \chi_{L}^c \rangle + \chi_{R}^T \tau^{\pm}_L\langle
\chi_{R} \rangle + \chi_{R}^{cT} \tau^{\pm}_L\langle \chi_{R}^c
\rangle + \nonumber \\ &  & \phi^T\tau^{\pm}_L\langle \phi \rangle
+ (\phi^c)^T\tau^{\pm}_L\langle \phi^c \rangle\bigg]{1 \over N_L}.
\end{eqnarray}
For definiteness let we focus on the negatively charged fields,
and let us work in the limit $u_1 \rightarrow 0$:
\begin{eqnarray}G^-_{L} &=& {1 \over \sqrt{u_2^2+w_L^2}}
\bigg[w_L\chi_{L}^{4,-{1 \over 2}} + u_2\phi^{-{1 \over 2},-{1
\over 2}}\bigg] \label{G1L}.\end{eqnarray} Likewise, $G^\pm_R$ are
associated with the SSB of charged sector in $SU(2)_R$:
\begin{eqnarray}G^\pm_{R} &=& \bigg[\chi_{L}^T
\tau^{\pm}_R \langle \chi_{L} \rangle + \chi_{L}^{cT} \tau^{\pm}_R
\langle \chi_{L}^c \rangle + \chi_{R}^T \tau^{\pm}_R\langle
\chi_{R} \rangle+ \chi_{R}^{cT} \tau^{\pm}_R\langle \chi_{R}^c
\rangle + \nonumber \\ & & \phi^T\tau^{\pm}_R\langle \phi \rangle
+ (\phi^c)^T\tau^{\pm}_R\langle \phi^c \rangle\bigg]{1 \over N_R},
\end{eqnarray} from which we obtain
\begin{eqnarray}
G^-_{R} &=& -{{1 \over
\sqrt{u_2^2+w_R^2}}}\bigg[w_R\chi_{R}^{4,-{1 \over 2}} -
u_2(\phi^{{1 \over 2},{1 \over 2}})^*\bigg].\label{G1R}
\end{eqnarray}
The states orthogonal to $G_{L,R}^\pm$ will be the physical Higgs
$H_{1,2}^\pm$ which in general not uniquely fixed because there
are 2 directions orthogonal to $G_{L,R}^\pm$.

A particularly important term in the Higgs potential is the term
responsible for mixing $\chi_L^{4,-{1 \over 2}}$ with
$\chi_R^{4,-{1 \over 2}}$ which effects the generation of Dirac
masses of neutrinos in the $\chi$-loop,
\begin{eqnarray}
V_3 = M \chi_L^{\dagger} \phi \tau_2 \chi_R  + M'\chi_L^\dagger
\phi^c \tau_2 \chi_R + \hbox { H.c.}. \label{V3}\end{eqnarray} We
will show that there exists an approximate global symmetry
$U(1)_X$ of this Higgs potential in the limit $u_1 \rightarrow 0,
\, u_2 \ne 0$ (or $u_2 \rightarrow 0, \, u_1 \ne 0$) that will
allow us to identify the physical Higgs states (in that limit).
The global symmetry $U(1)_X$ can be defined by the generator
\begin{eqnarray}
X = Y' + I_{3R} - I_{3L},
\end{eqnarray}
where \begin{eqnarray} Y' \langle \chi_R \rangle = -{1 \over 2}, \
\ Y' \langle \chi_L \rangle = {1 \over 2}, \ \ Y' \langle \phi
\rangle = 1. \end{eqnarray} Furthermore for the global symmetry
to be useful we require it to be unbroken by the vacuum, that
is,
\begin{eqnarray}
X \langle \chi_L \rangle = X \langle \chi_R \rangle = X \langle
\phi \rangle = 0
\end{eqnarray}
which is indeed the case given our choice of $X$ (and the limit
$u_1 \to 0$).

Referring to $V_3$ in Eq. (\ref{V3}), notice that the $M'$ term is
not a symmetry under $U(1)_X$. However, in the limit $u_1
\rightarrow 0$ then $M'$ must also be zero for self consistency.
The reason is that a non zero $M'$ in $V_3$ will induce a linear
term in the (21) component of $\phi$ (i.e. the place where $u_1$
would sit) when $\chi_{L,R}$ develop VEVs.  Because the potential
is linear in $u_1$  (for small $u_1$) a non-zero VEV for $u_1$
must arise which is obviously not self consistent with our
assumption that $u_1 = 0$. This shows that $U(1)_X$ becomes an
unbroken symmetry because the $U(1)_X$ asymmetry $M'$ term
vanishes in the limit $u_1 \rightarrow 0$. (We can also draw
similar conclusion in the limit $u_2 \rightarrow 0, \, u_1 \ne
0$.) Furthermore it allows us to uniquely specify the physical
Higgs fields $H_1, H_2$ since now we have two requirements. First
they must be orthogonal to $G_{L,R}$ and second they must be
composed of components with the same $U(1)_X$ charge. (Note that
$H_1$ has $X-$charge -1, $H_2$ has $X-$charge +1.) These
considerations lead to the identification of the charged physical
Higgs fields:
\begin{eqnarray}
H_1^- &=& {1 \over u_2\sqrt{1+\left(u_2/ w_R\right)^2}}
\bigg[{u_2^2 \over w_R} \chi_{R}^{4,-{1 \over 2}} + u_2(\phi^{{1
\over 2},{1 \over 2}})^*\bigg], \nonumber \\ H_2^- &=& -{1 \over
u_2\sqrt{1+\left(u_2 / w_L\right)^2}}\bigg[{u_2^2 \over w_L}
\chi_{L}^{4,-{1\over 2}} - u_2\phi^{-{1 \over 2},-{1 \over
2}}\bigg].\label{H12L}
\end{eqnarray}
Writing  Eq. (\ref{G1L}), (\ref{G1R}), (\ref{H12L}) in matrix
form, we obtain Eq. (\ref{lw1}). Note that the matrix $U_g$ is
unitary, $U_g^{-1} = U_g^T$ because the rows (and columns) are all
orthogonal. The existence of the approximate global symmetry $X$
in the limit $u_1 \rightarrow 0, u_2 \ne 0$ (or $u_2 \rightarrow
0, u_1 \ne 0$) means that the mass mixing term
\begin{equation}
V_{3\langle \phi \rangle} \propto u_1 u_2.
\end{equation}
The effect of this is that we can treat the mass mixing terms
$V_{3\langle \phi \rangle} = -(Mu_1 + M'u_2){\chi_L^{4,-{1 \over
2}} \chi_R^{4,-{1 \over 2}*}}$ as a small perturbation when $u_1$
$(u_2)$ is switched on from zero.

\end{document}